\documentclass[11pt]{article}
\usepackage[margin=1in]{geometry}
\usepackage{authblk}
\usepackage{times}
\usepackage[noadjust]{cite}
\usepackage[normalem]{ulem}
\usepackage[english]{babel}
\usepackage{blindtext}

\usepackage{booktabs}
\usepackage{url}
\usepackage{graphicx}
\usepackage{tikz}
\usepackage{amsmath}
\usepackage{epsfig,endnotes, graphicx,booktabs, multirow, array}
\usepackage{xspace}
\usepackage{colortbl}

\usepackage{tabularx}
\usepackage[utf8]{inputenc}
\usepackage{regexpatch}
\usepackage[font=footnotesize]{caption} %
\makeatletter
\newcommand{\linebreakand}{%
  \end{@IEEEauthorhalign}
  \hfill\mbox{}\par
  \mbox{}\hfill\begin{@IEEEauthorhalign}
}
\makeatother

\usepackage{algorithm}
\usepackage[noend]{algpseudocode}

\usepackage{subcaption}
\usepackage{xspace}
\usepackage{hyperref}
\usepackage[textsize=small,backgroundcolor=orange]{todonotes}
\usepackage{booktabs}
\usepackage{graphicx}
\usepackage{enumitem}
\usepackage{multirow,tabularx}
\newcommand{\html}{{HTML}\xspace}

\definecolor{aliceblue}{rgb}{0.94, 0.97, 1.0}
\newcommand{\name}{\mbox{\textit{CookieEnforcer}}\xspace}
\newcommand{\bertbasecased}{BERT\textsubscript{Base-Cased}\xspace}

\definecolor{aliceblue}{rgb}{0.94, 0.97, 1.0}

\newcommand{\set}{5k\xspace}

\usepackage{tcolorbox}
\newtcolorbox{mybox}[1]{colback=aliceblue,colframe=black,fonttitle=\bfseries,title=#1}

\hypersetup{
    colorlinks=true,
    linkcolor=blue,
    filecolor=magenta,      
    urlcolor=cyan
    }

\usepackage{cite}
\usepackage{breakurl}           %
\usepackage{url}                %
\usepackage{xcolor}             %
\usepackage[]{hyperref}         %
\hypersetup{                    %
  colorlinks,
  linkcolor={green!80!black},
  citecolor={red!70!black},
  urlcolor={blue!70!black}
}

\usepackage{mwe}
\usepackage{tikz}
\usetikzlibrary{arrows}
\usepackage{verbatim}

\usepackage{booktabs}
\usepackage{dirtree}
\usepackage{wrapfig}
\usepackage{caption}
\usepackage{subcaption}

\title{\Large \bf CookieEnforcer: Automated Cookie Notice Analysis and Enforcement}

\renewcommand*{\Affilfont}{\normalsize\normalfont}
\newsavebox\affbox

\author[1]{Rishabh Khandelwal}
\author[1]{Asmit Nayak}
\author[2]{Hamza Harkous\thanks{The opinions expressed in this publication are those of the author and not of Google.}}
\author[1]{Kassem Fawaz}
\affil[1]{%
  \savebox\affbox{\Affilfont Department of Chemical Engineering, University of AAAAA BBBBBB, CCCCC road,}%
  \parbox[t]{\wd\affbox}{\protect\centering} University of Wisconsin -- Madison} 
\affil[2]{Google Inc}
\date{}

\begin{document}

\date{}

\maketitle
\begin{abstract}
Online websites use cookie notices to elicit consent from the users, as required by recent privacy regulations like the GDPR and the CCPA. Prior work has shown that these notices use dark patterns to manipulate users into making website-friendly choices which put users' privacy at risk. 
In this work, we develop \name, a new system for automatically discovering cookie notices and deciding on the options that result in disabling all non-essential cookies. 
In order to achieve this, we first build an automatic cookie notice detector that utilizes the rendering pattern of the HTML elements to identify the cookie notices. Next, \name analyzes the cookie notices and predicts the set of actions required to disable all unnecessary cookies. This is done by modeling the problem as a sequence-to-sequence task, where the input is a machine-readable cookie notice and the output is the set of clicks to make.
We demonstrate the efficacy of \name via an end-to-end accuracy evaluation, showing that it can generate the required steps in 91\% of the cases. Via a user study, we show that \name can significantly reduce the user effort.
Finally, we use our system to perform several measurements on the top 5k websites from the Tranco list (as accessed from the US and the UK), drawing comparisons and observations at scale.
\end{abstract}

\section{Introduction}
\label{sec:introduction}

As a response to recent regulations from the EU and California, a cookie notice is almost a universal fixture in most websites. Devised initially to help websites maintain state on the user side, cookies have evolved to be a powerful tracking tool\cite{englehardt2015cookies, englehardt2016online, brookman2017cross, papadopoulos2019cookie}. Generally speaking, there are two types of cookies, the ones essential for the website operation, and the non-essential/unnecessary ones (used for enhancing the user experience or tracking their behavior). Cookie notices inform users about the type of cookies the website maintains, their purpose and, in many cases, the options to control them. However, in their current forms, cookie notices suffer from usability issues~\cite{habib2020s}. In some instances, these notices have become more cumbersome to interact with and are often distributed across multiple views. As we later show in this work, the users needed to click 12 times, on average, to disable non-essential cookies in our user study.

Take AccuWeather, when visited from the UK, as an example. To set their cookie settings, the user should first click on the ``Manage Options'' button to navigate to the settings menu. On that menu, there are 11 individual cookie settings, 9 of which are pre-enabled for ``legitimate interests." Further, there is another view for per-vendor settings with over 100 listed vendors along with their options; all these options are also pre-enabled. Both menus do not have an opt-out button; the user has to individually disable each cookie setting. Further, the cookie notice on this website does not allow the user to view the content unless they interact with the banner and give consent or adjust preferences. 

Therefore, users may find it hard to exercise informed cookie control for websites with complicated notices. They are far more likely to rely on default configurations than they are to fine-tune their cookie settings for each website~\cite{acquisti2006imagined, lai2006internet}. In several cases, these default settings are privacy-invasive and favor the service providers, which results in privacy risks\cite{liu2011analyzing,naini2015analyzing,privacy_scale:2004}. Several proposals have aimed at automating the interaction with cookie notices~\cite{consent_o_matic,ninja,cliqz}. These proposals, however, rely on manually analyzing a subset of cookie notices (by major Consent Management Platforms) and hard-coding JavaScript snippets to enforce privacy-respecting cookie choices. Such approaches do not scale with the breadth and depth of cookie notices. The implementation of cookie notices varies significantly across websites, calling for the need for a more scalable approach.

In this work, we propose a new paradigm to provide users with control over websites' cookie notices. We describe \name, a cookie enforcement controller system, which automatically finds cookie notices, locates the fine-grained options within these notices, understands the semantics of the cookies, and automatically disables non-essential cookies.
Achieving these objectives required (1) building a unified understanding of the cookie control settings that scales across web technologies and (2) identifying and automatically enforcing only the necessary cookies for each website.

\name utilizes machine learning techniques to discover and enforce cookie settings that protect the user's privacy. To address the challenges described above, \name leverages two key insights to enable the robust extraction and enforcement of cookie control elements: (1) their presentation to the user and behavior should be consistent to maintain the user experience, and (2) disabling unnecessary cookies via the cookie notice can be safely done without impacting the user experience.  Using both insights, \name applies a three-stage pipeline that, given a domain, extracts a machine-readable representation of its cookie controls and automatically disables unnecessary cookies.

First, \name crawls the domain and identifies the cookie notice via a machine-learning classifier that exploits the notice's textual and rendering features. Second, \name simulates users' behavior by interacting with every UI element on the notice. It extracts the controls within a notice and associates each control with its descriptive text and state. Third, we develop a decision model that takes in the text corresponding to all the cookie settings and their current state (selected or not-selected) to determine the actions required to disable the unnecessary cookies. We model this problem as a sequence-to-sequence learning task where the model maps the text and the state to the sequence of steps required. Towards that end, we finetune a deep text-to-text Transformer model (T5) model as the decision model~\cite{raffel_t5}. Finally, \name generates a JavaScript snippet that enforces the sequence of steps from the decision model for each domain. 

To turn \name into a usable tool, we built a Chrome browser extension that injects the generated JavaScript snippet within the \html of each visited domain. When the browser loads the \html, it will execute the snippet which disables all non-essential cookies. As such, \name automatically enforces the privacy-respecting settings for each cookie notice, without any further interaction, thereby making the cookie controls more accessible and reducing the overhead of the user at the same time.

We took several steps at different parts of building our system to motivate and evaluate the various design decisions. In particular:
\begin{itemize}[noitemsep]
    
    \item We make the case for a system like \name by conducting an online user study with 165 participants on Amazon MTurk, where we evaluate users' perception of cookie notices. We show the extent to which users are struggling to properly control their cookie settings.
    \item We further perform an end-to-end evaluation of \name over the top 500 websites from Tranco~\cite{le2019tranco}, assessing its core components. We show that our pipeline correctly generates a sequence of clicks required to disable non-essential cookies for 91\% of the pages in our manually annotated dataset. This evaluation showcases the generality of \name's design, despite the variance in the \html implementation of the analyzed pages and the dynamic flow of the notices.
    \item We further conduct an online user study with 180 participants on Amazon MTurk to measure the effectiveness of \name's client implemented as a browser extension. We show that it reduces the time taken to adjust cookie settings on a set of 14 popular websites by a factor of 2.35, without the need for any clicks. Moreover, \name obtained a 15\% higher score on System Usability Scale (SUS), compared to the manual baseline.
    \item Finally, we conduct measurements on the top-5k websites from the Tranco list, showcasing how \name can be used in the wild. Specifically, we find that 16.7\% of the websites with cookie notices when visited from the UK have at least one non-essential cookie enabled by default.
     
\end{itemize}

\section{Background and Related Work}
\label{sec:background}
Before we dive into the specifics of \name, we first provide the necessary background information about cookies, cookie notices and the HTML analysis techniques we use.
\subsection{Cookies}

A cookie is a small text file that a website stores in the users' browser. The original purpose of the cookies was to maintain user state in HTTP browsing, such as shopping carts and login sessions. Broadly speaking, cookies fall into two main categories: essential and non-essential. Essential cookies serve the core functions of the website, such as maintaining user sessions. Websites use non-essential cookies to perform additional operations, such as analyzing users' online behavior or providing targeted ads. 

Prior work demonstrated how cookies can enable tracking of the users' online activities~\cite{englehardt2015cookies, englehardt2016online, brookman2017cross, papadopoulos2019cookie}, posing significant privacy threats to web users. In response to these threats, recent regulations, such as the EU's GDPR and ePrivacy Directive 2009/1367EC, require websites to obtain consent from users before storing cookies. The GDPR  also states that consent must be freely-given, specific and informed. 

Cookie notices are the most widely adopted approach to meet these legal requirements; the websites usually show the cookie notices on the users' first visit. These notices  consist of interactive elements which allow users to set their preferences. Empirically, we observe that cookie notices usually have one or two views; the first view has coarse options like \textit{Accept} or \textit{Reject}. In several cases, a second view has fine-grained options to enable/disable cookies based on a per-purpose or per-vendor basis.  Consent Management Platforms (CMPs) help websites comply with these regulations~\cite{hils2020measuring}. These platforms are third party integrations, which provide easy solutions for obtaining and storing user consent. The adoption rate of these CMPs is still limited to 10\%  of the top 10,000 most popular websites~\cite{hils2020measuring}, with many websites opting to implement customized versions of the cookie notice.

\subsection{Cookie Notice Studies}

\subsubsection*{\textbf{Cookie Notice Analysis}}  Degeling et al.~\cite{degeling2018we} measured the GDPR's impact on cookie notices by manually examining the top 500 websites in each of the EU member states. They found that 62\% of the websites serve cookie notices. More recently, Kampanos et al.\cite{kampanos2021accept} used a list of common CSS selectors to detect cookie notices in 17000 websites in the UK and Greece. They found that 45\% of these websites serve a cookie notice. They also analyzed the notices to check for compliance to find that only a small fraction of websites provide direct opt-out option. Eijk et al.~\cite{eijk2019impact} used a similar methodology to understand the effect of user geo-location on the presence of cookie notices.  Matte el al.\cite{Matte:2020} compared the user options against those stored by the CMPs and found suspected violations. Bollinger el al.~\cite{bollinger2021analyzing} analyzed 30k websites and identified several GDPR violations. Finally, Coudert et al.~\cite{coudert2020automatically} used a keyword based scoring algorithm to detect cookie notices, and analyzed them for detecting dark patterns.

Our approach differs from these works in two aspects. First, we present a more robust cookie notice detection that does not rely on keywords or handcrafted rules (which can easily become obsolete). Second, we go beyond detecting cookie notices and extracting dark patterns. We analyze the detected cookie notices to extract and understand its fine-grained options using a deep text-to-text model. We use the understanding of these options to automatically disable non-essential cookies.

\subsubsection*{\textbf{Users' Perception and Dark Patterns}} Utz et al.\cite{Utz:2019} conducted a manual analysis to identify common properties of cookie notices. They investigated how these properties impact users' decision to accept/reject cookies, finding that nudging has a large effect on users' choice. Similarly, Machuletz et al. \cite{Machuletz2020} studied how does the number of options and presence of ``select all'' button influence users' decisions. Kulyk et al.~\cite{kulyk2018website} reported that users find cookie notices annoying and disruptive. Nouwens et al.~\cite{nouwens2020dark} studied the effect of CMPs on people's consent choices by scraping designs from popular CMPs in 10,000 UK websites, finding the presence of dark patterns on most of the websites. 

\subsubsection*{\textbf{Automated Enforcement}} The widespread availability of dark patterns in cookie notices motivated approaches for automated interactions on the user's behalf. Particularly, the browser extensions Consent-O-Matic~\cite{consent_o_matic}, Cliqz-Autoconsent~\cite{cliqz} and Ninja-Cookie~\cite{ninja} automatically enforce users' choices for cookie notices. However, these extensions employ rule-based detection and enforcement and rely on the presence of specific CMPs to function correctly. This approach does not scale to the majority of websites implementing customized cookie notices.  Similarly, other works~\cite{bollinger2021analyzing, hu2021cccc} classify cookies into pre-set categories and provide options to remove these cookies from the browser storage. In these approaches, the user is still required to interact with the cookie notices. \name addresses this limitation by emulating users' interaction with cookie notices.

Another set of works~\cite{khandelwal2021prisec, chendemystifying, habib2019empirical} analyze privacy settings pages to present them in a more accessible manner to the users. Specifically, Khandelwal el al.~\cite{khandelwal2021prisec} and Chen et al.~\cite{chendemystifying} automatically detect hard-to-find privacy settings on web and on android, respectively. Habib et al.~\cite{habib2019empirical} analyze the privacy policies of the websites to determine the opt-out links and presents them to the user. These approaches operate on fairly static webpages, and the user still has to manually interact with the settings. Our work differs in two aspects: First,we cope with the highly-dynamic nature of cookie notices. For example, in some cases, the cookie settings can be dynamically injected after the user interacts with the cookie notice (\textit{e.g.} clicks on ``More Options''). Second, these systems do not model the choices' semantics, whereas in \name, we use this modeling in order to (1) automatically disable the non-essential cookies, and (2) perform measurements around the websites' practices.

\subsection{HTML Analysis Techniques}

In order to detect the cookie notices, \name leverages techniques from the \html rendering process. HTML rendering can be abstracted as a relative ordering of layers (HTML elements) along an imaginary z-axis. The precise ordering of these layers, \textit{i.e.} which element is at the top and so on, is determined using the stacking context\footnote{For more details on stacking contexts: \url{https://developer.mozilla.org/docs/Web/CSS/CSS_Positioning/Understanding_z_index/The_stacking_context}.} and stacking order. The stacking order refers to the position of web elements on this imaginary z-axis. In the absence of special attributes, the stacking order is generally the same as the order of appearance in the HTML. This ordering can be altered using special CSS attribute called \textit{z-index}, where higher \textit{z-index} results in a higher position in the stacking order. The \textit{z-index} is set to ``auto'' for the elements where it is not specified explicitly.

\section{User Interaction with Cookie Notices} 
\label{sec:measurement_baseline}

Prior work has characterized user interaction with cookie notices, mostly focusing on users' acceptance rate for different configurations~\cite{Utz:2019}. However, the user effort required to adjust cookie settings has not been studied, despite being a common pain point\cite{hofstad2021cookies, kulyk2018website}. To motivate the design of \name, we first perform a manual analysis on 100 popular websites to estimate the user effort required to disable non-essential cookies. We then conduct an online study to explore the factors that affect the users' decisions.

\subsection{User Effort Required to Disable Cookies}
We manually examined the top-100 websites from Tranco's~\cite{le2019tranco} most popular website list from a UK-based IP address (through a VPN service). We visited each website using the Selenium\footnote{\url{https://selenium.dev}} library, with ChromeDriver\footnote{\url{https://chromedriver.chromium.org}}. One of the authors manually disabled the non-essential cookies (wherever possible) and recorded the required number of clicks; we use the number of clicks as a proxy to user effort. 

We find that, on the 48 websites (with English content) that showed a cookie notice, the user has to perform an average of 3.5 clicks to disable non-essential cookies. Further, we note that 13\% of the websites did not have option to disable non-essential cookies, whereas 17\% of them had a one-click option to disable non-essential cookies. The maximum number of clicks required was 19. Note that we are reporting a lower bound on the number of required clicks as the author is a domain expert.

\subsection{Online User Study}
Next, we conduct an online user study to understand user preferences towards the cookie settings and explore the factors that affect their decision. We develop this study using partial disclosure by hiding its purpose from the participants. 

\paragraph*{\textbf{Study Flow}} 

We ask each participant to visit four websites (from a pool of 14 websites) after verifying that their browser do not have cookies for these destinations (thus ensuring that they see the cookie notice). Then we ask them to answer two questions related to content of the website. This procedure ensures that the participants interact with the website and the cookie notice as they normally would, without any priming or instructions around the cookie notice itself. 

In the study, the participants first install a monitoring extension that we deployed on the Google Chrome Webstore. Then the plugin identifies the websites from our set which do not have cookies stored in the participants' browser. From these websites, we randomly select four websites and ask the user to visit the websites. That way, we ensure that the user sees cookie notices on all websites. We note here that the participant is asked to disable Ad-blockers for the duration of the study as some of them block cookie notices.

After the user finishes their tasks, we ask them about their familiarity with browser cookies and cookie notices. We also ask them about their general preference for browser cookies and finish the study with a qualitative question asking them to explain any factors that influence their decision to allow/reject cookies. These questions are consistent with the qualitative study conducted by Hofstad et al.~\cite{hofstad2021cookies}.  We did not ask for any personally identifiable information, and the IRB at our institute approved the study. Additional details on the user study, including snapshots, can be found in Appendix~\ref{sec:appendix_baseline_study}.

\paragraph*{\textbf{Website Selection}} 
As our primary objective in this study is to understand user preferences towards cookie notices, it is necessary that each participant sees the cookie notice on all the selected websites. Therefore, to minimize overlap with users' browsing history, we purposefully select a set of 14 non-popular websites (the list is in Appendix~\ref{sec:appendix_baseline_study}).

\paragraph*{\textbf{Participant Recruitment}} 
We recruited {161} participants from Amazon Mechanical Turk. We choose participants with $>90\%$ HIT approval rate to control for quality of the participants. Further, we restricted the location of participants to be in the United states. We paid each participant {$\$2.75$} to participate in the study that lasted {13} minutes on average. Of these {165} participants, 67.68\% were males, 32.32\% were females; 69.09\% of the participants had a Bachelors or equivalent four-year degree and 10.30\% had a graduate degree. The average age range of the participants was between 25-34.

\subsection{Study Findings}	
\label{sec:baseline_res}
Here, we first discuss the various factors affecting users decision. We then analyze the users' preferences with respect to the cookie notice.
 
\subsubsection{Coding Qualitative Responses} We asked the participants to qualitatively describe the factors that influence their decision with respect to cookie settings. For a systematic analysis of the responses, two of the authors manually coded the responses into four high-level categories. The coders exhibited a high agreement in this categorization (Cohen's $\kappa = 0.85$)\cite{landis1977measurement}. They had disagreements in 7 cases, which they were able to resolve after discussions.  

The first, and most popular, category was ``Forced interaction'' where the users interact with the cookie notice because they are required to do so (e.g., for websites with blocking cookies). The second category was ``Risk," where the users factor the risk associated with accepting cookies into their decision. The third category was ``Misinformed,'' where exhibit wrong understanding of the cookies. The last category was ``Trust,'' where the participants' trust in the websites affects their cookie decisions. We provide example quotes from each category in Appendix~\ref{sec:appendix_baseline_study}.

\begin{figure}[t]
  \centering
  \includegraphics[width=\columnwidth]{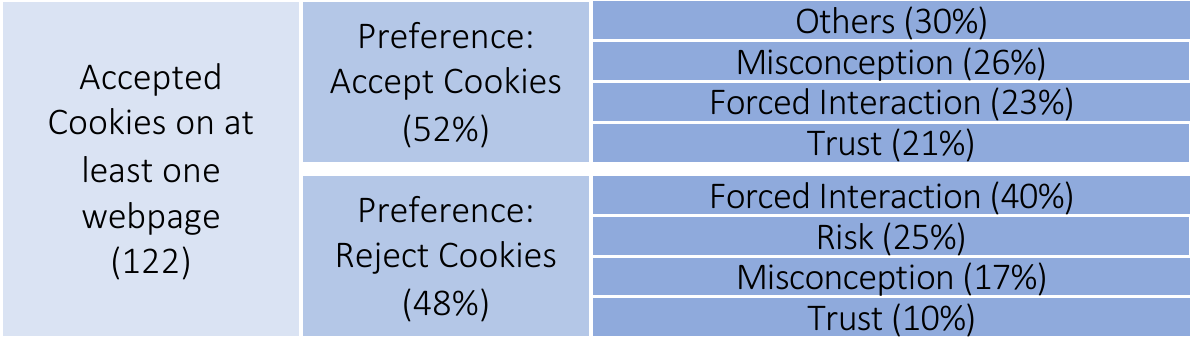}
  \caption{Analysis of participants who accepted cookies on at least one webpage.}
  \label{fig:comment_analysis}
\end{figure}

\subsubsection{Analyzing Users' Behavior}
We analyze how users interacted with the cookie notices and compare their behavior with their answers in the post-study questionnaire. First, we find that 12\% of the users did not interact with the cookie notices, 12\% rejected cookies when they interacted, and 76\% accepted cookies from at least one website. 

Fig.~\ref{fig:comment_analysis} shows the breakdown of responses from participants who accepted cookies in at least one website. We observe that 48\% of these participants indicated that they do not want to allow cookies, yet they do. The coded answers revealed that most of these participants allow cookies (against their stated preference) because of the forced interaction nature of the notice as in the case of one participant: ``\textit{I prefer not to be tracked, but sometimes there’s no choice.}''

The rest of these participants (52\%) had a different distribution of reasons as to why they want to allow cookies. Most of these reasons centered around misconceptions about the nature of cookies. For instance, one participant mentions that ``\textit{I think cookies are necessary as it helps a system to load that particular website more fast the next time an user accesses the website.}'' The other leading reason was the forced interaction nature of the notice. For example, one participant mentioned that ``\textit{If im visiting a site for the first time and I cant reject cookies, i usually accept it}''. Interestingly, none of these participants mentioned risk as a factor in their decision.

\subsubsection{Takeaways}
Cookie notices often interrupt the user flow while they are surfing on the web. We find that a user needs to execute an average of 3.5 clicks to disable non-essential cookies on top-100 websites. In the user study, we find that more than half (53\%) of the users either did not accept non-essential cookies or indicated that they did not want to accept non-essential cookies. Furthermore, 30\% of the participants mentioned that being forced to interact with the cookie notice affected their behavior. These findings, combined with the observation from Hofstad et al.~\cite{hofstad2021cookies} about users being annoyed and concerned for their privacy by the cookie notices, motivate the use case for an automated system to disable non-essential cookies.

In this work, we provide such a solution with \name. The goal of \name is to automatically find cookie notices, understand them, and disable non-essential cookies. In the next section, we start with an overview of the different components of \name.
Over the following sections, we discuss the design and implementation of these components.

\section{System Overview}
\label{sec:overview}

\begin{figure}[t]
    \centering
  \includegraphics[width=\columnwidth]{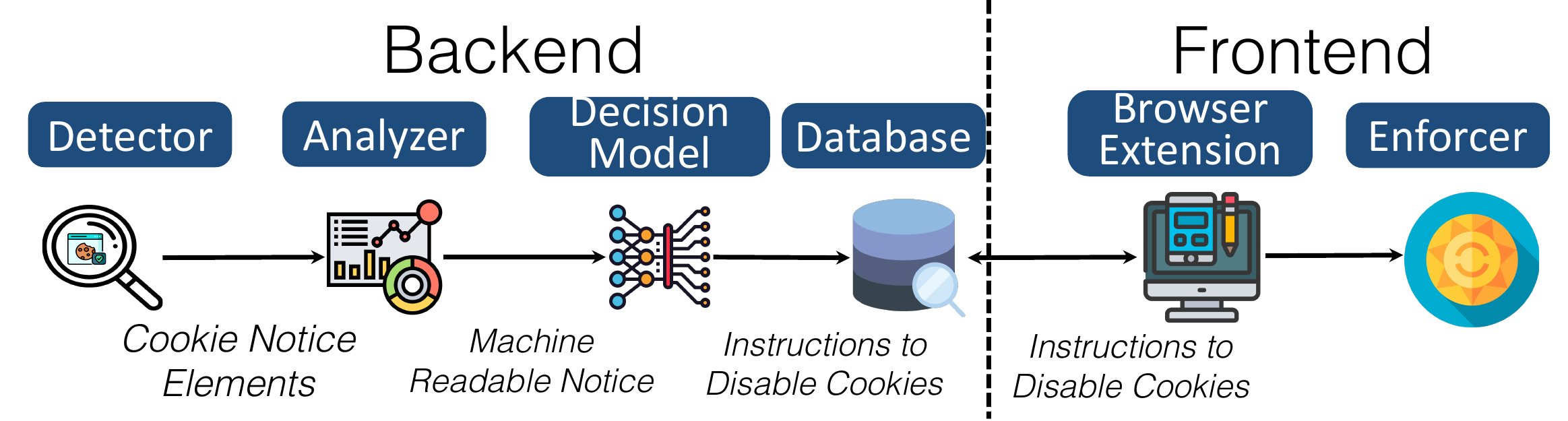}
    \caption{An overview of \name's components. \textit{Backend} generates the machine readable representation of cookie notices whereas \textit{Frontend} uses them to disable non-essential cookies.}
    \label{fig:overview}
\end{figure}

The three objectives of \name are to transform the cookie notices into a machine readable format, determine the cookie setting configuration to disable non-essential cookies (whenever possible), and automatically enforce this configuration. A high level overview of \name is in Fig.~\ref{fig:overview}; it utilizes two components to achieve its objectives.
\begin{itemize}
    \item The \textit{backend} component is responsible for detecting and analyzing the cookie notices. This component generates the necessary steps to disable non-essential cookies.

\item The \textit{frontend} component consists of a browser extension that uses the instructions from the backend to generate and execute the JavaScript code that automatically disables non-essential cookies.
\end{itemize}

\subsection{Backend}
The \textit{backend} of \name consists of three modules. The \textit{Detector} module (Section~\ref{sec:detector}) takes as input a domain name and identifies the web element corresponding to a cookie notice (if present). Then, the \textit{Analyzer} module (Section~\ref{sec:analyzer}) mimics the behavior of a human user by dynamically interacting (performing click actions) with the cookie notice to locate all the adjustable settings. 
This module accounts for the cases where settings become unhidden or are dynamically injected upon user interaction. It outputs a list of all interactive elements and their associated text description. Next, the \textit{Decision Model} (Section~\ref{sec:decision}) utilizes semantic text understanding to determine the settings configuration (the values for each preference), which disables the non-essential cookies. Based on the identified configuration, the backend generates the steps required to perform automatic enforcement. We implemented the backend component in \name, using the Selenium library with ChromeDriver. Selenium automates the interaction with the cookie notice to extract the relevant settings.

\subsection{Frontend}
The \textit{frontend} consists of the \name browser extension which fetches the information for each website from the \textit{backend} and generates the required JavaScript code to disable the cookies. Locally, the extension determines whether the cookie notice is visible for a given website. If the notice is found, the plugin injects the necessary JavaScript to disable the non-necessary cookies. Note that the cookie notice might not appear if the cookie settings have been decided on before (by the user or the extension).

\subsection{Challenges}

In order to achieve the goals of \name, we must overcome four main challenges:
\begin{itemize}
    \item First, \name must identify the cookie notice present on the website. This problem is challenging due to the flexible nature of \html implementation. For example, prior work~\cite{eijk2019impact} that used CSS selectors to detect cookie notices had a high false negative rate of 18\%.
    
    \item Second, \name must extract the configurable settings along with their context from the cookie notice. This task is complicated by the dynamic nature of HTML. For example, interactable elements can be dynamically injected in the notice using JavaScript, making the static analysis of \html ineffective. 

    \item Third, to represent a cookie notice in a machine readable format, \name needs to identify the various effects an element can have once it is clicked, in an automated manner. For example, in Fig.~\ref{fig:element_explain}(b), \textit{Save Settings} button submits user preferences whereas the switch disables/enables cookies.

    \item Finally, \name must understand the context of each cookie setting. This task is also challenging since the context of the settings (provided by the text describing them) comes from free form natural language, and is diverse. Keyword-based approaches cannot scale with the diversity of text in cookie notices. For example, on \url{www.virginmedia.com}, the element that reveals fine-grained settings has the text: \textit{``Open the cookie jar''}.
\end{itemize}

\section{Cookie Notice Detector}
\label{sec:detector}
The \textit{Detector} module detects the presence of cookie notices on webpages. It analyzes the \html from a webpage and returns the web element corresponding to the cookie notice, if present. As indicated earlier, this  task is challenging as the open nature of \html allows different implementations of the cookie notices. For example, it is possible to design the cookie notices as floating pop-ups with custom elements, inline frames (\textit{IFrames}), \textit{shadow-roots},\footnote{For more details:  \url{https://developer.mozilla.org/docs/Web/Web_Components/Using_shadow_DOM}} or simply as \textit{div} elements. \name addresses these challenges by relying on the global stacking order of \html. First, it retrieves a set of candidate notice elements. Second, it extracts the text from these elements. Finally, it uses a text classifier to determine whether the element is a cookie notice.

\subsection{Candidate identification}

A website serving a cookie notice is expected to surface the notice as the user lands on the webpage. As such, the elements corresponding to the cookie notices should be higher in the stacking orderof the \html.\footnote{Technically, cookie notices should be higher in the stacking order of the \html within the root stacking context. We omit references to the stacking context for simplicity.}
As described in Section~\ref{sec:background}, the stacking order determines which element the user sees on the top most layer of the webpage. The \textit{Detector} module leverages this invariant behavior. It looks for a non-negative \textit{z-index} attribute within the stacking context and tags them as potential candidate elements.  However, in practice, not all implementations of cookie notices utilize the z-index to surface the cookie notices. For example, the website \url{www.gov.uk} shows the notice as the first element in the \html tree. To capture such instances, the \textit{Detector} module tags the first and the last three visible elements of the webpage as candidates.

\subsection{Text Classifier}
After obtaining the candidates, our goal is to identify the cookie notice element. We rely on the text in the candidate elements and use a text classifier to perform this task. 

One approach to perform this classification is to use a keyword-based model as the cookie notice is expected to convey information about the use of cookies. However, this approach is not effective for cases which provide notice and choice without explicitly talking about the cookies. For example, when accessed from the United Kingdom, the cookie notice on \url{www.arizona.edu} reads: \textit{I have read, understand, and consent to UA's enrollment management Privacy Policy. Consent, Decline}. Therefore, we need a classification model that relies on the text semantics to determine if the candidate element is a cookie notice. 
 
To this end, we use a text classifier based on BERT (Bidirectional Encoder Representations from Transformers), which is transformer based encoder system pretrained on masked language modeling and next sentence prediction objectives~\cite{devlin2018bert}. BERT has been the model of choice for achieving strong performance on a variety of text classification tasks, such as sentiment analysis or topic classification~\cite{sun2019fine}. The key advantage of a large pretrained model like BERT is that it is readily trained on a large corpus, so it can be finetuned on a downstream task with a relatively small dataset. In this work, we finetune the \textit{\bertbasecased} (case-sensitive variant with 12 layers) to determine whether a given candidate element is a cookie notice.

\subsubsection{Training and Performance}
We create the data for the classifier by sampling 250 websites from the top-50k most popular website list from Tranco~\cite{le2019tranco}. We first extract the candidate elements for each website from this set by using the candidate identification methods. One of the authors then manually annotated each website, indicating whether it showed a cookie notice. 
The annotation task involved looking at the screenshots of the webpages and identifying if a cookie notice was present. As the task is fairly easy for an expert, we only require one annotation per website. We obtain 112 websites with cookie notices and 138 without cookie notices. We extract at most two candidate elements from each website to obtain a total of 505 candidate elements, 112 of which are notice elements. From this set, we keep aside a balanced test set of 100 candidates, 50 cookie notices elements and 50 non-cookie notice elements.

For each candidate, we first extract its text by concatenating the text of all the elements inside it. For example, in Fig.~\ref{fig:element_explain}(a), the input text for the classifier would be: 

\noindent\textit{We use cookies to improve your browsing experience...to manage your cookie settings, click ``More Information''. \textit{Accept Cookies} \textit{More Information}.} 

Next, we train the classifier on the training set with 62 notice elements and 343 non-notice elements. We use oversampling during training to ensure that both classes were represented equally. We trained the \bertbasecased model with a learning rate of $2e^{-5}$ for 10 epochs and used the last model checkpoint for evaluation. Table~\ref{tab:cookie_classifier} shows the performance of the classifier on the test set. The classifier achieves an average F1-score of 0.97, indicating that the model learned to distinguish cookie notice elements from the rest. Analyzing the failure cases, we observe that, in a few cases where the text contained topics other than cookies, the model was confused. 
We attribute this to the fact that as text about other topics increase, the information about cookie notices present in the text gets diluted, resulting in mis-classification.

\begin{table}[t]
\footnotesize
\caption{A breakdown of the classifier's performance on the test set.}
\centering
\begin{tabular}{lcccc}
\toprule
\textbf{Instances} & \textbf{Support} & \textbf{Recall} & \textbf{Precision} &\textbf{ F1-score }\\
\midrule
\rowcolor{aliceblue}Not Cookie notice  & 50 & 0.96 & 0.98 & 0.97\\
Cookie notice & 50 & 0.98 & 0.96 & 0.97\\
\midrule
\rowcolor{aliceblue}Total Pages & 100 & 0.97 & 0.97 & 0.97\\
\bottomrule
\end{tabular}

\label{tab:cookie_classifier}
\end{table}

\section{Cookie Notice Analyzer}
\label{sec:analyzer}

The \textit{Analyzer} module takes the HTML element corresponding to the cookie notice as its input and extracts the cookie settings, 
their current state (selected or not-selected), and the text corresponding to the settings. The analyzer passes a list of the extracted cookie settings to the Decision Model (Section~\ref{sec:decision}). The latter enables \name to determine the configuration, which disables non-essential cookies\footnote{Video with steps can be found here: \url{https://youtu.be/ViyKxbY3rAM}}.

The flexible nature of HTML implementations presents two challenges for the \textit{Analyzer} module. First, cookie notices are frequently dynamic. On several websites, the elements corresponding to cookie settings only load when another button is clicked. This issues renders the static analysis of HTML ineffective. Second,  the fine-grained cookie settings in many of the cookie notices are initially hidden. In order to change the fine-grained settings, users have to navigate to a different view (usually by clicking buttons like ``Cookie Settings''). This second view is usually a different element in the DOM tree. As a result, \name has to keep track of the browser state with respect to the different cookie elements as well as different view of the cookie notice.

\name addresses these challenges by mimicking the actions of real users: it interacts with the cookie notices and observes the behavior of the webpage after each interation. The \textit{Analyzer} starts by first discovering the elements in the notice with which the user can interact. Here, it leverages the \textit{tabbing} feature provided by the \html which allows the user to access interactable elements by pressing the \texttt{Tab} key. Next, the \textit{Analyzer} clicks on each element to identify any dynamically injected elements.
Finally, it identifies the cookie settings and extracts the text corresponding to those settings.

\subsection{Identifying Interactive Elements}
\name leverages the \textit{tabbing} feature of \html to identify the interactive elements within the cookie notice. This feature was originally introduced to enhance the accessibility and reach of webpages by allowing users to access interactive elements via the \textit{Tab} key. Prior work, analyzing the \html pages to detect privacy settings, also used this technique~\cite{khandelwal2021prisec}. The key idea is that, since the users need to interact with the cookie settings to adjust the preferences, we can simulate this interaction via \textit{tabbing} and obtain a set of candidates for cookie settings. We further supplement this set by adding hidden \textit{input, button} and \textit{anchor link} elements. By relying on this invariant behavior of the \html, \name extracts the set of candidate cookie settings. 

The set of candidates obtained from the tabbing do not contain dynamically injected elements.
Dynamically injected elements are loaded as a result of an interaction with another element. For example, in Fig.~\ref{fig:element_explain}, the settings appearing after clicking on ``More Information'' button are dynamically loaded.  The \textit{Analyzer} module recursively checks for these elements by clicking each visible element from the candidate set and querying again to find new elements.

After obtaining the candidate elements set, the \textit{Analyzer} module excludes the elements that redirect the user to a different page or open a separate tab. This way, we filter out links for cookie policies, explanations about cookies and cookie vendor details. 
A side effect of this decision is that the module also filters out elements which take user to dedicated webpages for cookie settings. For example, \url{linkedin.com} (when accessed from the UK), contains an option which leads to a dedicated page for cookie settings. We discuss the implications of this decision in Section~\ref{sec:discussion}.

\subsection{Extracting Cookie Settings}

\begin{figure}[t]
  \centering
  \includegraphics[width=\columnwidth]{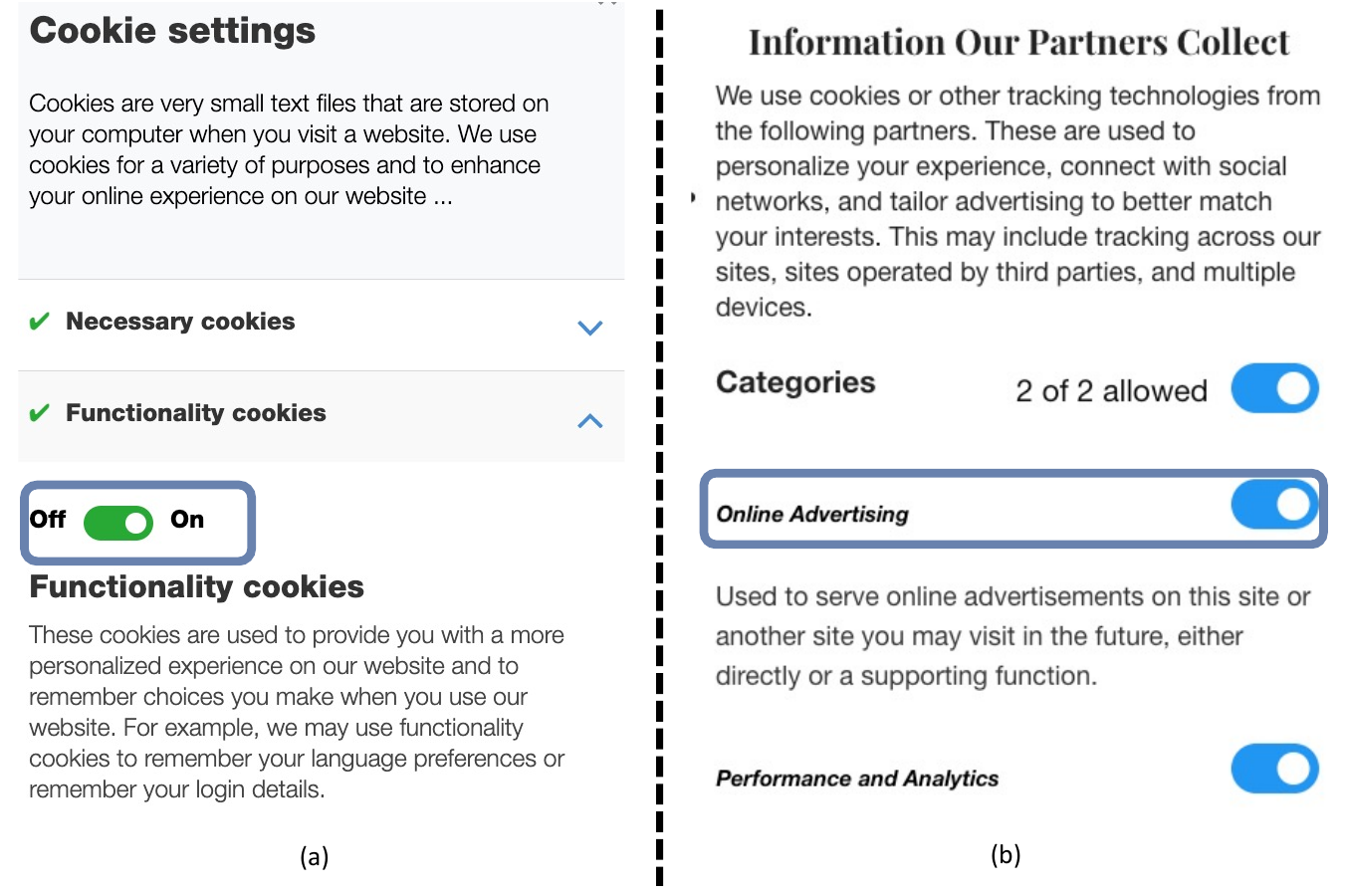}
  \caption{Examples of different types of text extraction. (a)Switch on \url{www.horiba.com} has no \textit{aria-label}, text is extracted via \html code and on-screen distance. (b) The label for switch on \url{www.justinbeaber.com} has \textit{aria-label} as Online Advertising}
  \label{fig:text_extraction_helper}
\end{figure}

At this point, we assume that the analyzer found all interactable elements in the cookie notice.
The next step is to extract the text that describes these settings. This text, combined with the state of the element (selected/not-selected) is needed for the decision model (Section~\ref{sec:decision}) to semantically understand the cookie notice.

Here, we use two independent signals to extract descriptive and concise text corresponding to an \html element. First, we leverage the \textit{aria-label} attribute,\footnote{For more information: \url{https://www.w3.org/TR/wai-aria/\#aria-label}.} wherever available. This attribute allows assisted technologies to read and consume webpages, thereby making web content accessible to users with disabilities. For example, the \textit{aria-label} attribute for the highlighted switch in Fig.~\ref{fig:text_extraction_helper}(b) has a value of ``Online Advertising'' which describe what setting the switch adjusts. 

In the absence of \textit{aria-label} attribute, we design a text extraction technique inspired by Khandelwal et al.~\cite{khandelwal2021prisec}. 
This technique extracts the text which provides details about a given element. For each interactable element, it searches for the closest parent node in the DOM tree that contains text. However, this parent node might contain other text such as the description of the setting.
For example, in Fig.~\ref{fig:text_extraction_helper}(a), ideally we would like the text corresponding to the switch to be \textit{Functionality cookies}, as opposed to ``Functionality cookies'' together with the description below it. We address this limitation by relying on the on-screen distance to identify the element describing the setting. Specifically, we find the closest (on-screen) text containing element from the cookie setting. For example, in Fig.~\ref{fig:text_extraction_helper}(a), the closest text element for the switch (marked with the box) is \textit{Functionality cookies}.

\section{Decision Model} %
\label{sec:decision}

At this stage, we have extracted all the interactable cookie settings and their corresponding text. 
The next step is to represent the cookie notice and its settings in a machine readable format, and determine the actions required to disable non-essential cookies. This is done in two steps. First, we understand the effect of each element as it is clicked, \textit{i.e.,} we determine the execution role for each element. Execution roles capture the various effects elements can have upon interaction with them. For example, the element tagged as (A) in Fig.~\ref{fig:element_explain} allows the user to enable/disable a particular cookie. Next, we understand the context (as provided by the setting text and the execution role) in which the user interacts with the settings. We then use the contexts of all the settings to determine the configuration required to disable the non-essential cookies.

\subsection{Determining Execution Roles}
\label{sec:execution_roles}

\begin{figure}[t]
  \centering
  \includegraphics[width=\columnwidth]{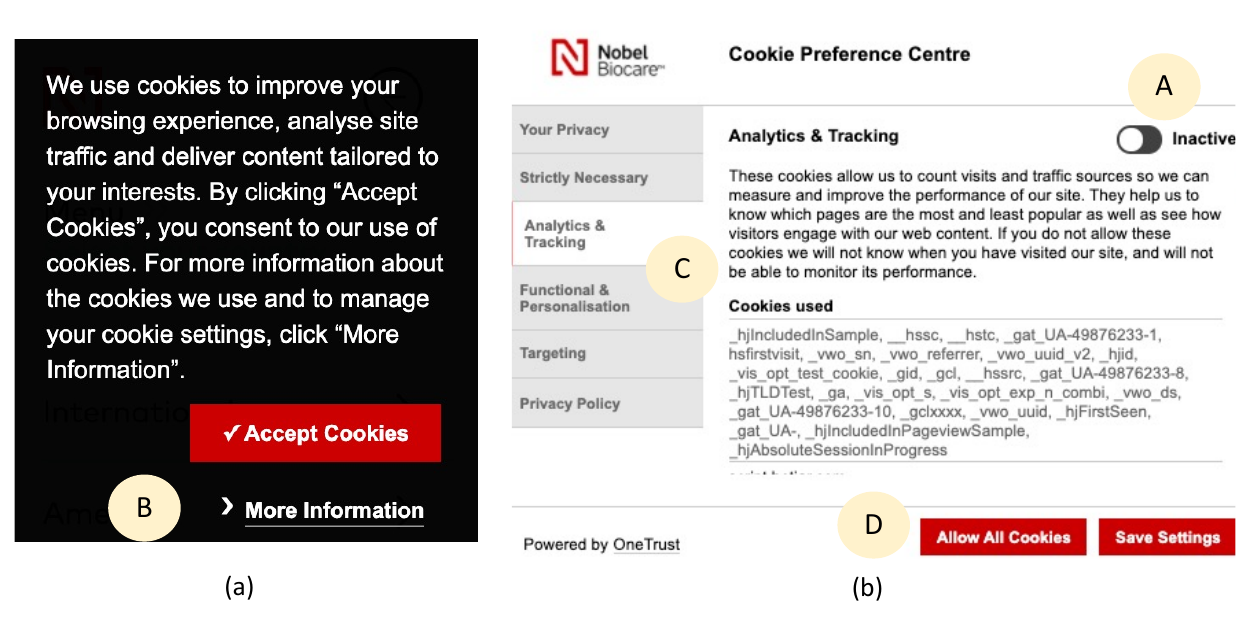}
  \caption{Cookie notices from \url{www.nobelbiocare.com} showing elements with different execution roles. (A) Type A element used to enable/disable Analytics and Tracking cookie. (B) Type B element that reveals the second banner shown on the right. (C) Type C element revealing the hidden settings. (D) Type D element to submit the preferences.}
  \label{fig:element_explain}
\end{figure}

In order to represent a cookie notice in a machine readable format, \name 
determines the execution role of the elements by interacting (performing the click action) with them and analyzing the effect on the webpage.
We define the execution role for all interactive elements within the cookie notice as described in Table~\ref{tab:execution_roles}. These roles are a result of categorizing some of the possible outcomes when the user clicks an element. Type A elements allow a user to adjust their preference for a particular setting. For example, the switch element A in Fig.~\ref{fig:element_explain} enables/disables Analytics cookies. Type B elements reveal new cookie notices. For example, in Fig.~\ref{fig:element_explain}(a), the button tagged B (``More Information'') opens the notice element shown in Fig.~\ref{fig:element_explain}(b). Similarly,  Type C elements reveal hidden settings within a cookie notice (e.g., ``Functional and Personalization'' tab in Fig.~\ref{fig:element_explain}). Finally, Type D elements are used to submit the choices. Examples are the buttons saying ``Allow All Cookies'' or ``Save Settings''. Type D elements typically conclude the users' interaction with the cookie notice.

\subsubsection*{Background on clicking actions in Selenium} To provide an intuition of how to identify execution roles, we first give a brief background on clicking actions in Selenium. There, the click command emulates the click operation on a given element. However,
the click action can only be performed if the element is visible on the browser (and is not overlayed by another element). For instance, if the first click on the \textit{Save Settings} button removes the notice, a second click will result in an error as the element is no longer visible. Another example is when an element allows users to configure a choice. There, we should be able to click it multiple times to change the choice. We leverage these behaviors to identify the different execution roles for the elements.

To identify the execution roles of the elements, the \textit{Analyzer} module clicks on each element twice (with a small delay between clicks) and checks if the element is still visible. Specifically, it uses the following criteria to determine the execution roles:
\subsubsection*{\textbf{Type A}} An element belongs to Type A if it is visible after two clicks and its state (selected or not-selected) changes with the clicks. For example, the switch element in Fig.~\ref{fig:element_explain} changes states and is visible after the clicks.
Note that it is possible to implement Type A elements such that the state cannot be queried; however, empirically, we found that to be very rare.

\subsubsection*{\textbf{Type B}} Elements belonging to Type B reveal another cookie notice. Thus, to identify these elements, we check (1) if the element disappears after the clicks, and (2) the \textit{Detector} module returns the new notice. For example, when we click the button ``More information'' in Fig.~\ref{fig:element_explain}, the new notice (the right plot in Fig.~\ref{fig:element_explain}) appears. Thus, we determine the execution role of the button to be Type B.

\subsubsection*{\textbf{Type C}} To identify Type C elements, we require that (1) the element can be clicked twice, and (2) that its \textit{checked} attribute should not change with clicks. These elements are used for internal navigation within the notice. 

\subsubsection*{\textbf{Type D}}  Such elements result in closing the cookie notice. We identify these elements by requiring (1) failure in the second click, and (2) no new cookie notice appearing after first click.

\begin{table}[t]
\footnotesize
\caption{Definition of the execution roles with examples.}

\begin{tabular}{m{0.8cm} 
>{\arraybackslash} m{4cm}
>{\arraybackslash}m{10.0cm}}
\toprule
\textbf{Type} & \textbf{Execution Role} & \textbf{Example}                                       \\ \midrule
\rowcolor{aliceblue} 
A             & Configuring choices    & A switch enabling/disabling marketing cookies \\ \midrule
B & Uncovering hidden notices  & Cookie Settings button in Fig.~\ref{fig:element_explain} (B) that reveals another notice when it is clicked        \\ \midrule
\rowcolor{aliceblue} 
C & Uncovering hidden settings & Analytics and Tracking Cookies tab in Fig.~\ref{fig:element_explain} (C) that reveals setting which was previously not visible \\ \midrule
D & Enforcing choices      & Accept Button in Fig.~\ref{fig:element_explain} (D) that completes the users' interaction with the notice.         \\ \bottomrule
\end{tabular}

\label{tab:execution_roles}
\end{table}

\subsection{Need For Natural Language Understanding}
\label{sec:nlu}
At this stage, we have extracted all the cookie settings from the cookie notice, and determined their execution roles. Next, \name needs to parse the different settings across all views of the cookie notice and semantically understand them.

One approach to perform this task is to simply deselect all the enabled options and determine which element to click to save the configuration. However, this approach has two main limitations. First, the existing settings are not always enabled or disabled by default. The user might be required to interfere to enable/disable cookies. The website \url{www.microsoft.com} has one such example for the Analytics, Social Media and Advertising cookies. Second, the cookie setting might be worded in a way where the element needs to be selected to disable non-essential cookies. For example, the option can be: \textit{Only allow necessary cookies}. Deselecting this option will lead to undesirable outcomes. Hence, it is important to account for the text of the element too.

Another approach is to treat the action associated with each interactable element as an independent classification problem where the task is as follows: given the text associated with the element, determine if the element should be clicked. The major drawback with this approach is that it models the task as a series of decisions without considering the interplay between these decisions. For example, take a website that has the following options: ``Disable Analytics Cookies,'' ``Accept Cookies,'' and ``Save Configuration.'' In this case, the model needs to know whether Analytics cookies are disabled before deciding whether to click on ``Save Configuration''. Therefore, the decision about whether to click an element cannot be made without the information about the current state for other elements. 

Thus, we observe that an effective decision model should meet two requirements: a) semantically understand the text corresponding to the options, and b) determine the series of actions required by accounting for all the options. 

\subsection{Extracting Actions to Disable Cookies}

Our main goal here is to develop a system which takes in the text corresponding to all the cookie settings and their current state (selected or not-selected), and determines the actions required to disable the non-essential cookies. Keeping up with the requirements identified earlier, we model this problem as a sequence-to-sequence learning task where the model gets the text and the state and determines the steps required. Specifically, we train a Text-To-Text Transfer Transformer (T5) model as the decision model. 

The T5 model, introduced by Raffel et al.~\cite{raffel_t5}, proposes a unified framework that treats all NLP tasks as text-to-text problems. This model has been shown to have a strong performance on a variety of NLP tasks ranging from classification to generation problems. T5 was also successfully applied to problems where the output is not necessarily traditional prose text. For instance, it has been applied to the text-to-SQL generation problem~\cite{shaw-etal-2021-compositional,furrer2020compositional}, where the output is a code to be executed.
The general approach of serializing structured steps into text has also been used to achieve state-of-the-art results in the data-to-text generation community~\cite{harkous-etal-2020-text,kale-rastogi-2020-text}.

T5 was released with multiple pre-trained checkpoints, ranging from ``small''
(60 million parameters) to ``11B'' (11 billion parameters).
For our purposes, we fine-tune a T5-Large model (770 million parameters) to produce a sequence of steps (clicks) required to disable the non-essential cookies. We first transform the information stored about the cookie notice in a single sentence format. This transformation is guided by the execution roles (Table~\ref{tab:execution_roles}) of the elements. Specifically, Type A elements have a state associated with them (selected/not-selected) whereas the other elements do not. The state of Type A elements allows the model to understand that these elements are configurable. Then we train the model to produce a text indicating which elements to click, given the text representation. 
The input and output for the T5 model would take the following format:

\begin{mybox}{\footnotesize \centering Input-Output format for the Decision model.}
\raggedright	
\noindent{\scriptsize \textsf{\textbf{Input}: \mbox{$<$ {notice\_0\_tag\_0}$>$ - $<${notice\_0\_tag\_0\_text}$>$}, \mbox{$<${notice\_0\_tag\_0\_state}$>$ $||$ $<${notice\_0\_tag\_1}$>$ - $<${notice\_0\_tag\_1\_text}$>$}, \mbox{$<${notice\_0\_tag\_1\_state}$>$ - \ldots} **
\mbox{$<${notice\_1\_tag\_0}$>$ - $<${notice\_1\_tag\_0\_text}$>$ - \ldots}
 $<$end$>$}}

\noindent{\scriptsize\textsf{\textbf{Output}: 
\raggedright
Click $<${notice\_0\_tag\_0}$>$ $|$ 
Click $<${notice\_0\_tag\_2}$>$
** 
Click $<${notice\_1\_tag\_2}$>$
}}
\end{mybox}
The ** symbol separates multiple notices' contents in the input and the output.
The $||$ symbol separates the settings options within the same notice in the input. The  $|$ symbol separates the click steps within the same notice in the output. Note that the state for an element is only defined if it belongs to Type A. For example, the input and output for the T5 model corresponding to the cookie notices on \url{www.askubuntu.com} shown on (1) and (4) in Fig.~\ref{fig:user_interface} are presented below. 

\begin{table*}[t]
\caption{Examples demonstrating the application of \textit{Decision model} on cookie notices for a few websites. We show the screenshots corresponding to these cookie notices in Fig.~\ref{fig:cookie_examples} (in Appendix~\ref{sec:appendix_t5_examples}) Note that for \url{www.tata.com}, the options are non-standard but the decision model is still able to reject the cookies.
}
\begin{center}
\begin{tabularx}{\textwidth}{m{2.3cm} m{10cm} m{3cm}}
\toprule
\textbf{Website} & \textbf{Input} & \textbf{Output}  \\
\midrule

\rowcolor{aliceblue} \url{www.reddit.com} & button0 - reject non-essential $||$ button1 - accept all  $<$end$>$ & Click button0. \\

 \url{www.netflix.com} & button1 - learn more about our use of cookies and information. $||$ button4 - accept $||$ button5 - reject $||$ button6 - personalise my choices $||$ button7 - close ** button0 - close $||$ switch5 - advertising cookies , not selected $||$button27 - save settings  $<$end$>$ & Click button5. \\

\rowcolor{aliceblue}\url{www.wordpress.com} & button0 - customize $||$ button1 - accept all ** switch3 - analytics: these cookies allow us to optimize performance by collecting , selected $||$ switch4 - advertising: these cookies are set by us and our advertising , not selected $||$ button5 - accept selection  $<$end$>$ & Click button0 ** Click  switch3 $|$ Click  button5. \\

\url{www.tata.com} & button0 - sweet! $||$ button1 - sorry, i'm on a diet $<$end$>$ & Click button1. \\

\rowcolor{aliceblue}\url{www.newscientist.com} & button1 - i accept $||$ button2 - show purposes ** button4 - select basic ads; object to legitimate interests ... switch23 - analytics cookies , not selected $||$ button62 - confirm my choices & Click button2 ** ... Click button4 $|$ Click button62. \\

\bottomrule
\end{tabularx}
    
\end{center}
\label{table:t5_examples}
\end{table*}

\begin{mybox}{\footnotesize \centering Example of a training sample.}
\noindent{\scriptsize \textsf{\textbf{Input} : button0 - customize settings $||$ button1 - accept all cookies ** switch3 - performance cookies, not selected $||$ switch4 - functional cookies, not selected $||$ switch5 - targeting cookies, not selected $||$ button6 - confirm my choices $||$ button7 - accept all cookies $||$ button8 - cancel $<$end$>$ }}\newline
{\scriptsize \textsf{\textbf{Output} : Click button0 **   Click button5. }}
\end{mybox}

We note here that some websites provide an option to opt-out of non-essential cookies on the first cookie notice but can have pre-selected options on the second. When creating the training data, we chose to disable the cookies from the first notice only. This emulates the behavior of the human who would not click to see more options if the option to reject non-essential cookies was provided. The model learns this behavior too upon training. This way, the decision model, given all the options available on a given webpage, can predict what actions to take to disable non-essential cookies.

\subsection{Training and Performance}
\label{sec:dec_training}
To create the dataset for the decision model, we first sample 300 websites with cookie notices from Tranco's top-50k popular website list~\cite{le2019tranco}. Next, we analyze the sites using the \textit{Detector} and the \textit{Analyzer} module to extract the options and their states (selected or not-selected). Then, one of the authors manually determined the series of clicks required to disable the non-essential cookies. This resulted in a dataset of 300 labeled websites. Next, we keep 60 websites aside for the test set. We further ensure that the test set has diverse instances across the types of cookie notices. Next, with the remaining data, we train a T5-Large with a batch size of 16 for 20 epochs with a learning rate of 0.003. For this task, we set the maximum input sequence length as 256 tokens and the maximum target sequence length as 64. These tokens are sub-words determined by the SentencePiece tokenizer~\cite{kudo2018sentencepiece}.

To test the performance of the model, we measure its accuracy on the test set. Note that the accuracy metric used here is the exact match percentage: the generated sequence should be exactly same as the ground truth. However, in practice, this restriction can be relaxed depending on the output sequence. For example, the relative order of clicking on two switches is often not important, but clicking the ``Save'' button before clicking a switch might give undesirable outcomes. Here, we take the conservative approach, and use the exact match percentage as the metric. We find the accuracy of the model on the test set to be 95\%, indicating that the model has succeeded in learning the task across a variety of websites. For example, given the input: \\
\noindent{\small\textsf{\textbf{Input} : switch0 - do not allow non-essential cookies, not selected $||$ button1 - save $||$ button2 - accept $<$end$>$ }}\newline
The model correctly generates:\\
{\small\textsf{\textbf{Output} : Click switch0 $|$ Click button1. }} \newline
We note that this phrase was not present in the training set. The most similar phrase to this was: \textit{do not sell personal information}. 

Table~\ref{table:t5_examples} shows examples from applying the decision model on a diverse set of cookie notices (the screenshots for these notices are shown in Fig.~\ref{fig:cookie_examples} of Appendix~\ref{sec:appendix_t5_examples}). Notably, we see that for \url{www.netflix.com}, there are two views for the cookie notice with second view consisting of fine grained options. However, since the first view contains a \textit{reject} button, the decision model only clicks on it. Another interesting example is \url{www.newscientist.com}. We have truncated the input due to space constraints. Apart from the regular switches, the second view for cookie notice on this website contains an option to \textit{object to legitimate interests} for basic ads. This option can be easily missed by the users as they have to expand an additional frame to see that. \name not only finds this option, but also understands the semantics and decides to object. These examples showcase that the model learns the context and generalizes to new examples.
We further evaluate the performance of the decision model with a larger dataset in the evaluation (Section~\ref{sec:evaluation}).

Finally, \name stores all the extracted information in a database and makes it available for the \textit{Frontend}. This information contains instructions on how to reach a cookie notice and interact with the desired elements to disable non-essential cookies. Prior work~\cite{gunawan2019comparison} has used XML Path Language (\textit{XPath})~\cite{clark1999xml} to reference the \html elements. However, we empirically found that, due to the dynamic nature of the notices, \textit{XPaths} for cookie notices are highly vulnerable to change upon page updates (\textit{e.g.} in the DOM tree, notice element can be injected before or after another div element is loaded for ads); hence they are not suitable. Instead, we rely on the \texttt{querySelector()} HTML function\footnote{For more details: \url{https://developer.mozilla.org/en-US/docs/Web/API/Document/querySelector}} (which returns the element matching a specified CSS selector or group of selectors in the \html). Using this function, we construct a path that can be used to identify the elements, even when the placement of the element is dynamic.

\section{Frontend}
\label{sec:frontend}

\begin{figure}[t]
    \centering
    \includegraphics[width=0.75\columnwidth]{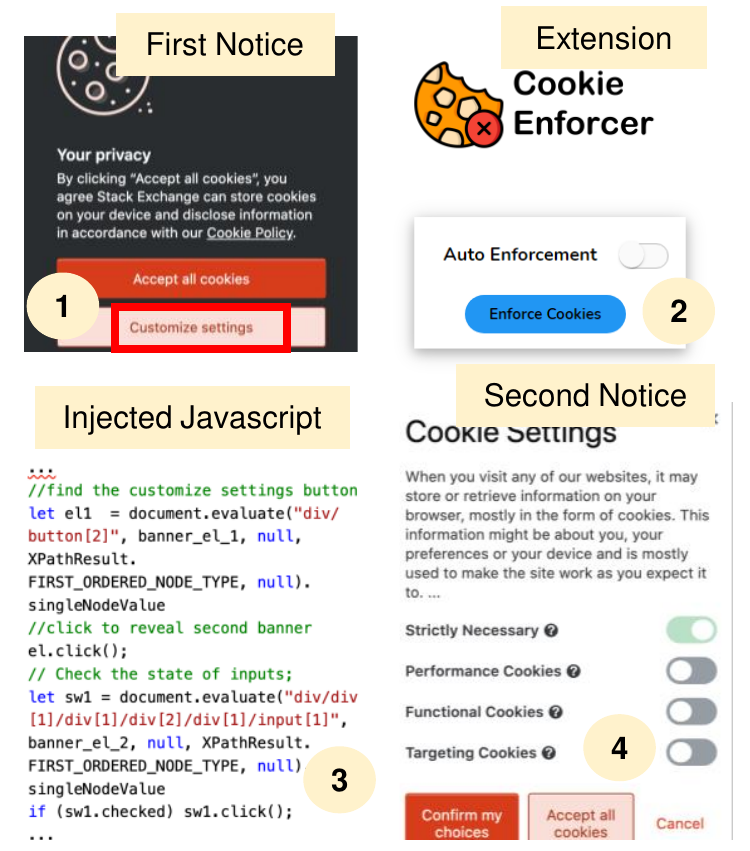}
    \caption{A typical workflow of \name extension with manual enforcement. (1) First the user visits \url{www.askubuntu.com}. (2) User activates the plugin and instructs the extension to disable non-essential cookies. (3) \name retrieves the information (locally) and generates the Javascript required. (4) Adjusted settings before submitting preferences.}
    \label{fig:user_interface}
\end{figure}

The frontend of the \name is a browser extension for Google Chrome. 
The extension periodically retrieves the set of all supported domains from the backend along with the set of instructions required to disable non-essential cookies for a each domain. 
This setup avoids exposing the user to a privacy risk; it does not involve sending each visited website to the backend for receiving the instructions.
Using the instructions, the extension generates  and injects the JavaScript required to disable the non-essential cookies. 
The main components of the frontend are the user interface and the \textit{Enforcer} module. 

\subsection{User Interface}

As one of the goals of \name is to provide automated enforcement to disable non-essential cookies, the user interface only consists of one switch and a button. The switch provides users' the option to enable/disable automated enforcement on all websites they visit, whereas the button is used to trigger enforcement when automated enforcement is not activated. This design decision is motivated by the results of the qualitative analysis done on users' responses in Section~\ref{sec:baseline_res}. There, we observed that some users enabled non-essential cookies based on websites' trustworthiness and utility. This switch allows the users to trigger manual enforcement on selected websites.

With automated enforcement, the extension retrieves the instructions to disable non-essential cookies from the offline data. The instructions contain the \texttt{CSS selector} path for the cookie notice extracted by the \textit{Detector} module~\ref{sec:detector}. Next, using the \texttt{CSS selector} path as input to \texttt{querySelector()} \html function, it determines whether the cookie notice is present. Note that the notice may not appear if the user has already set the preference. After detecting the cookie notice, it triggers the \textit{Enforcer} module (discussed below). On the other hand, with manual enforcement, the process only starts after the \textit{Enforce Cookies} button is clicked.

\subsection{Enforcer}

This \textit{Enforcer} module takes in the set of instructions from the user interface and generates the JavaScript required to disable the non-essential cookies. Fig.~\ref{fig:user_interface} shows a typical workflow of the plugin on \url{www.askubuntu.com} for manual enforcement. The user visits the webpage and sees the cookie notice. Next, the user activates the plugin and chooses to enforce cookie settings. The \textit{Enforcer} module then generates and injects the Javascript, which checks for the current states of the inputs and configures the required states for them. A working demo can be found here: \url{https://youtu.be/5NI6Q981quc}

\section{Evaluation}
\label{sec:evaluation}

We perform experiments to evaluate \name and showcase its utility in large scale analysis. Specifically, we answer the following questions:

\begin{itemize}

    \item\textbf{Q1. \textit{What is the end-to-end performance of \name?}}
    
We perform an end to end evaluation of \name on 500 domains that it has not seen before.  Our evaluation shows that \name generates correct instructions to disable non-essential cookies for 91\% of the websites.
    
    \item\textbf{Q2. \textit{Does \name improve user experience with cookie notices?}}
    
         We evaluate the usability of \name by conducting an online user study with a set of 14 websites. We measure the System Usability Scale (score) from user feedback and compare \name against a manual baseline.  We find \name obtains 15\% higher score than the baseline. Further, \name also reduces the time taken to disable cookies by 58\%.
    \item\textbf{Q3. \textit{Can \name analyze cookie notices in the wild?}}
    
Powered by the semantic understanding abilities of \name, we analyze the nature of options provided by cookie notices in the top 5k websites in the US and Europe. While measurements at such a scale have been performed before, \name allows a deeper understanding of the options beyond keyword-based heuristics. In particular, we find that 16.7\% of the websites in the UK showing cookie notices have enabled at least one non-essential cookie. The same number for websites in the US is 22\%.
    
\end{itemize}

\subsection{End to End Evaluation}
\begin{figure}[t]
    \centering
    \includegraphics[width=\columnwidth,scale=0.9]{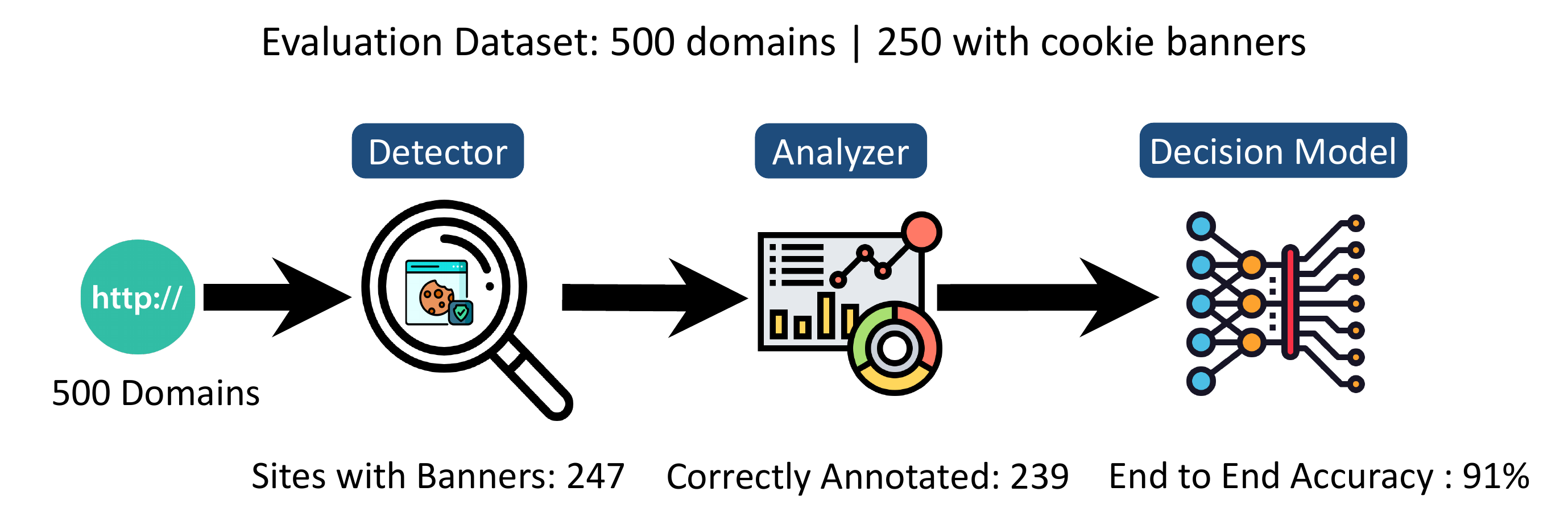}
    \caption{The results from evaluation of \name shows that the system performs well on the test set.}
    \label{fig:end_to_end_eval}
\end{figure}

We perform an end-to-end evaluation of \name on 500 websites. The goal is to take in domains that the system has not seen before and extract a machine-readable representation of cookie notice (if present). We then manually verify the correctness of this process. This evaluation consists of evaluating the individual components of \name, namely, the \textit{Detector} module (Sec.~\ref{sec:detector}), the \textit{Analyzer} module (Sec.~\ref{sec:analyzer}) and the Decision Model (Sec.~\ref{sec:decision}). 
We show a high level overview of these steps in Fig.~\ref{fig:end_to_end_eval}

\subsubsection{Dataset}
For the end-to-end evaluation, we sample a balanced evaluation set of 500 websites from the top-50k websites as ranked in the Tranco list~\cite{le2019tranco}. We then manually annotate the dataset for the \textit{Detector} module by taking a screenshot of the webpage and determining if a cookie notice is present. For the \textit{Analyzer} module, manually annotating the dataset to extract the cookie settings beforehand is not feasible as there is no unique identifier for these options. Thus, we manually verify the existence of all the cookie settings after we pass the data through the \textit{Analyzer}. To create the annotated dataset for \textit{Decision Model}, we obtain the input string for the model from the \textit{Analyzer module} and determine the steps required to disable non-essential cookies based on the string.
As we want to evaluate \name on a diverse set while still representing the top websites, the evaluation set contained 250 domains from the top-1k and 250 domains in the 1k to 50k range of the Tranco list. Further, as the EU region is expected to show more cookie notices, we perform the evaluation by accessing the websites from the United Kingdom via a VPN.

\subsubsection{Findings}
We first pass the 500 domains of the evaluation set through the \textit{Detector} module. The module tagged 247 domains as having cookie notices. This set contained 246 domains that have cookie notices (out of the 250 tagged in our manual curation). Only one website had a cookie notice falsely detected (an element with a link to the cookie notice).

Two of the domains that the \textit{Detector} module missed were due to the websites detecting the usage of an automated tool, thus restricting access. On the third missed website, the cookie notice is only shown for about 6 seconds before it disappears, so it was missed by our tool as it has an included delay to allow all elements to load. Interestingly, that notice's default setting is to enable all cookies.
The last website that the module missed was because the cookie notice was present under a special element, called the ``shadow-root''\footnote{For more details:  \url{https://developer.mozilla.org/docs/Web/Web_Components/Using_shadow_DOM}}. This element allows the encapsulation of other content (similar to an \textit{IFrame}), but it is rendered separately from the document's main DOM tree. We could not easily access such elements via an automated browser. 

At this stage, we have 247 domains tagged as having a cookie notice element (including one false positive). Next, we pass these notice elements to the \textit{Analyzer} module which identifies the various cookie settings present in the notice. Upon manual verification, we find that the \textit{Analyzer} module correctly identifies the options in 97\% of the websites. When performing manual verification, we count a website as an error if the \textit{Analyzer} misses at least one cookie setting.
We further analyze the cases that the \textit{Analyzer} module missed and find that, in most cases, the options we missed had a non-standard \html implementation.
Specifically, we missed three cases because the interactive elements were not reachable via the tabbing feature. 
One of the websites we missed here had a bug in its notice implementation preventing it from disappearing after clicking on ``Accept'' or ``Reject''. 
Moreover, we note here the \textit{Analyzer} module also filters out the single false positive case from the previous stage as the element on that page only had out-of-page links.

Finally, we pass the elements from the remaining 239 domains to the Decision Model. We use the model to generate the outputs and compare the generated sequences with the manually annotated ones. We use the exact sequence match as the metric here (as discussed in Section~\ref{sec:dec_training}). Here, we find that the decision model accurately predicts the correct steps 94.8\% of the time. The majority of the failed instances were domains where the number of settings was too high due to which the input to the model got truncated\footnote{The T5 model we finetuned allows using up to 512 input sequences while we were using only 256 at inference time.}. 

Thus, even with the variations in the \html and the dynamic nature of elements in the cookie notice, our end-to-end pipeline accurately generates the steps required to disable cookies in 227/250 websites, thereby achieving an overall accuracy of 91\%.

\subsection{User-based Evaluation}
\label{sec:eval_user_study}
We evaluate the usability of \name by conducting a user study on Amazon Mechanical Turk. This study is a separate from the one conducted in Section~\ref{sec:measurement_baseline}. We recruited 180 participants from within the United States with $>90 \%$ HIT approval rate. Among our recruited participants, 35\% were female, 65\% were male and 78\% had at least a Bachelor's degree. The average age range was found to be 25-34 years. We paid \$2.25 for the study, which lasted for an average of 11 minutes and 50 seconds with a median time of 9 minutes and 50 seconds. We did not ask for any personally identifiable information, and the IRB at our institute approved the study. 

\subsubsection{Study Design}
We instructed each participant to visit two websites and disable all non-essential cookies. 
We followed a \textit{within-subject} design, where the same user was exposed to two conditions.
In the first condition, the user had access to the automated cookie enforcement option in \name. In the second (baseline) condition, the user had to manually find the settings and disable the non-essential cookies.  We randomized the order of these conditions to account for learning and fatigue effects. 
We note here that the goal of this study was to measure usability of the extension. Hence, we did not obscure the goal of disabling cookies in the experiment.

\paragraph*{\textbf{Website Selection}} To measure the usability, we require that the participants are able to see and interact with the cookie notices. Here, we use the same initial pool of websites from the study in Section~\ref{sec:measurement_baseline}. We purposefully selected a set of 14 non-popular websites (complete list is in the Appendix~\ref{sec:appendix_baseline_study}) to minimize the overlap with users' browsing history.

During the study, all participants first install the \name extension from the Google Chrome Webstore. Then the plugin identifies the websites from our set, which do not have cookies stored in the participants' browser. From these websites, we randomly select two websites and ask the user to disable the non-essential cookies, using the plugin on one website and using the baseline (manual) method on the other. Note here that the order of conditions (baseline vs plugin) is randomized, as discussed above. In the manual condition, the participant interacts with the cookie notice to disable the non-essential cookies whereas, in the plugin condition, we instruct the participants to load the page, click on the extension icon, and use the \textit{Enforce Cookies} button (pane 2 in Fig.~\ref{fig:user_interface}) to complete the task.

As the participants interact with the cookie notice, we record the total time, and the number of clicks it takes for them to adjust the cookie settings. We define the start time as the time when the website has loaded in the browser (measured via the extension). We also monitor the elements that they click, which are used to determine the end time for the task.  After each task, the participants fill the System Usability Score questionnaire\cite{brooke1996sus}. Finally, there is an open ended question asking for general feedback at the end of the survey. More details about the user study are included in the Appendix.~\ref{sec:appendix_baseline_study}.

\begin{figure} 
    \centering
  \subfloat[System Usability Scale\label{fig:SUS}]{%
        \includegraphics[width=0.5\columnwidth]{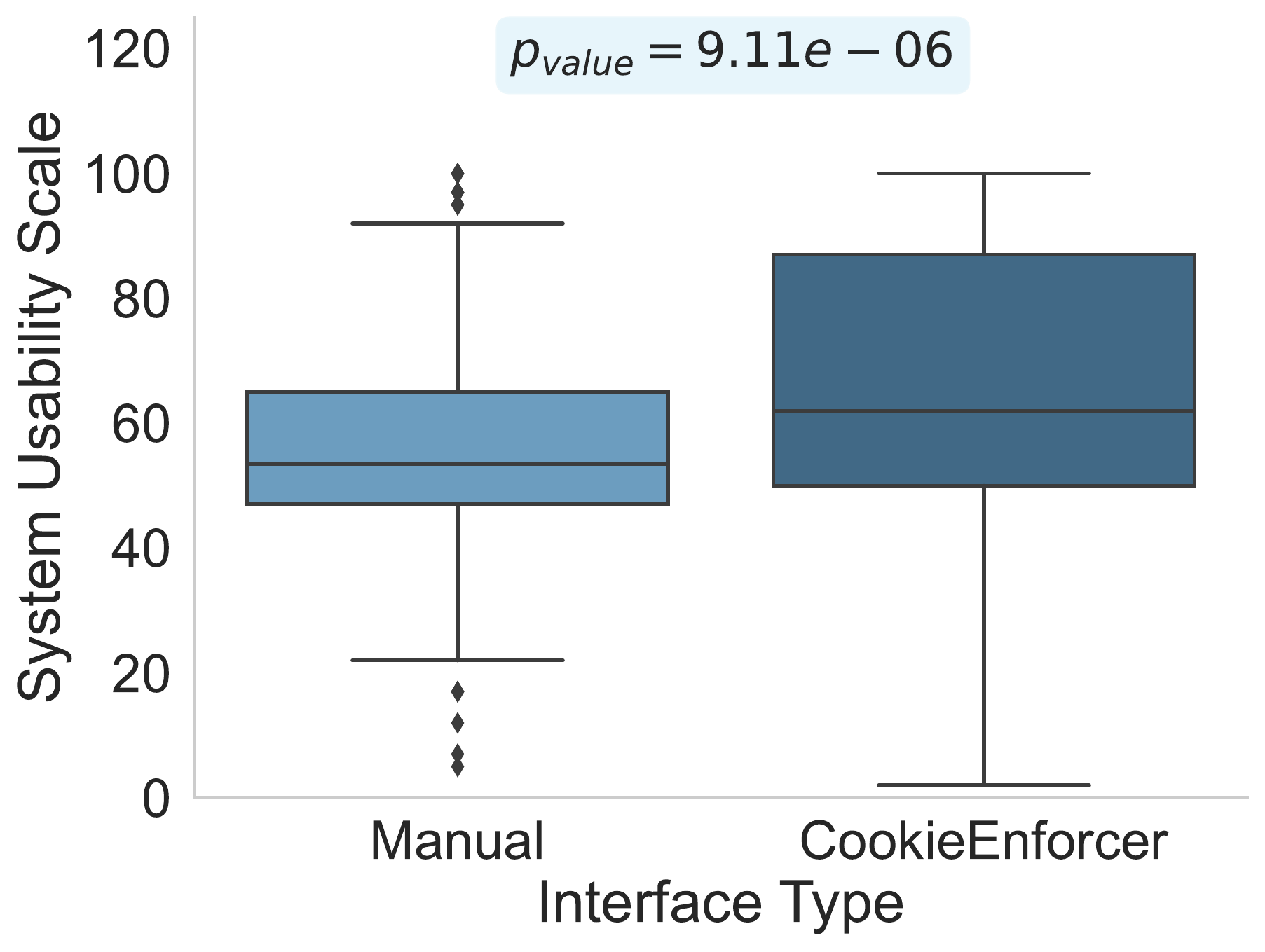}}
    \hfill
  \subfloat[Time per task\label{fig:Plugin}]{%
      \includegraphics[width=0.5\columnwidth]{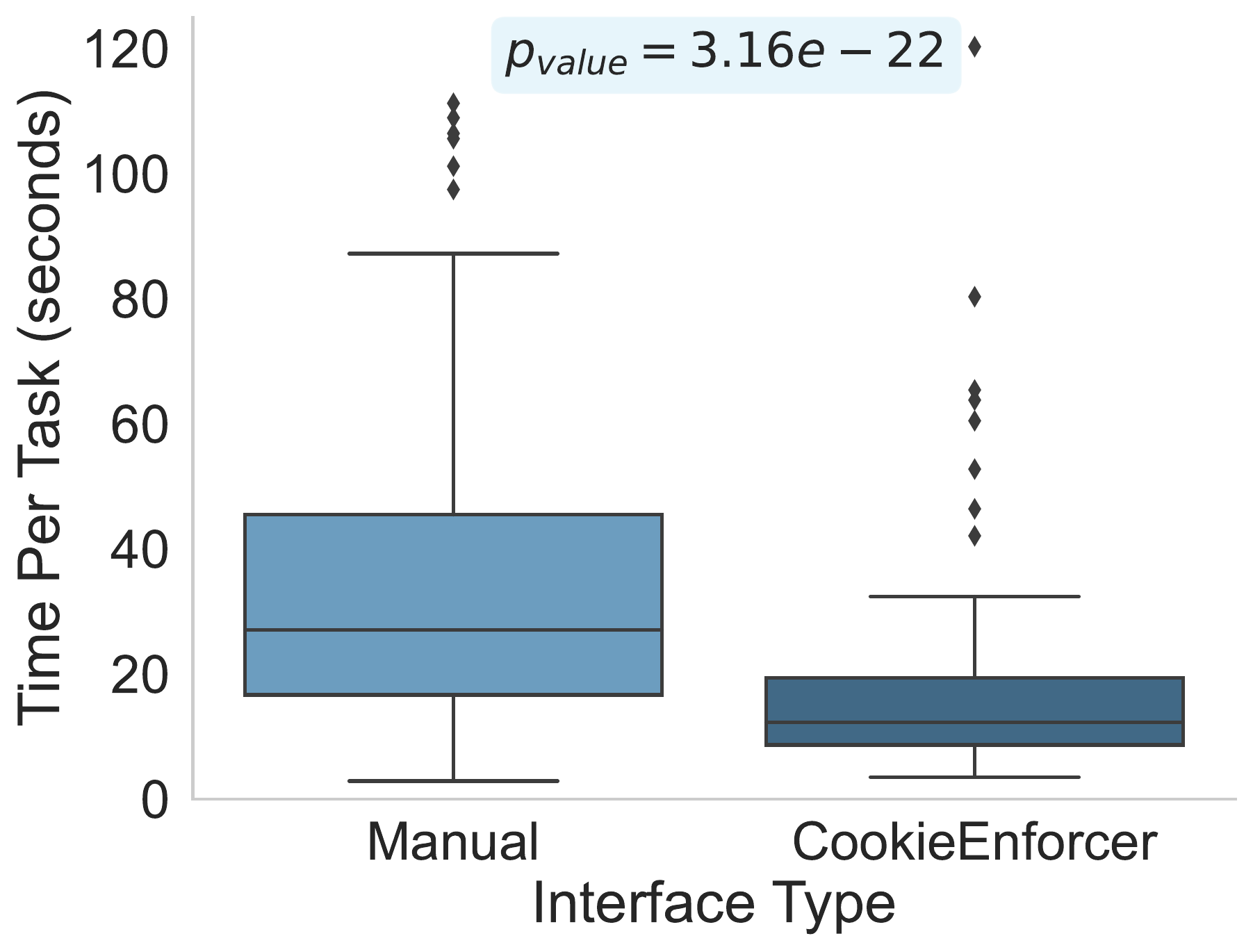}}
  \caption{The results from the usability study; (a) shows an increase in usability of \name versus the manual baseline, (b) shows the decrease in time taken by users to adjust cookie settings using \name. }
  \label{fig:SUS_Plugin} 
\end{figure}

\subsubsection{Findings}
\label{sec:eval_user_study_findings}

We assess the usability of \name using two metrics: 1) usability score as measured by the SUS survey and 2) user effort as measured by time taken by participants to disable non-essential cookies.

Fig.~\ref{fig:SUS} compares the System Usability Scale (SUS) score for the baseline system (manual) and the \name plugin. SUS scores have been used in the literature to evaluate different UI designs~\cite{bangor2008empirical}. In our study, we find that the \name obtained a 15\% higher score on System Usability Scale (SUS), compared to the manual baseline. We test the statistical significance of this change in score using 
using the Wilcoxon signed-rank test~\cite{wilcoxon1992individual}. We find that the result is statistically significant, and we reject the null hypothesis with a \textit{p-value} of $9.1\times 10^{-6}$. 

Next, we compare the average time taken by the participants to complete the task using the \name plugin and baseline system in Fig.~\ref{fig:Plugin}. It is noteworthy that during automated enforcement, we add a delay of 1 second after each click (to give ample time for the clicks to take effect). We report the results including this 1 second delay.
We find that, on average, users needed 13.57 seconds to complete the task with the \name plugin, whereas they needed 32.08 seconds with the baseline system. This shows that the \name plugin reduces time taken to adjust preferences by a factor of 58\%. 
We again test the statistical significance using the Wilcoxon signed-rank test~\cite{wilcoxon1992individual} and reject the null hypothesis with a \textit{p-value} of $2.5\times 10^{-3}$. Additionally, we also note that average number of clicks required to finish the task in the baseline (manual) system was 12 clicks per site, whereas with the \name plugin, the user would not be interrupted as the enforcement can be done automatically as the page loads. Thus, the \name plugin significantly reduces time taken by the user while disabling non-essential cookies on the cookie notice.

\subsection{Measurement}
\label{sec:measurement}

Finally, we demonstrate how \name can be used to study cookie notices in the wild. We first describe the dataset that we used, and then discuss the methodology and results.

\subsubsection{Dataset}
We consider the top-5k websites from Tranco's most popular website list. We filter out {1082} non-english websites using the \textit{langdetect} library~\cite{langdetect}. Additionally, we also filter out 344 websites which we were not able to access using the automated browser. 
At the end of this filtering, we have 3574 websites on which we perform our measurement.

\subsubsection{Methodology}
We first pass the websites through the \textit{backend} of the \name and obtain a machine readable representation of the cookie notices and the series of actions required to disable non-essential cookies wherever possible. Using these, we measure the following quantities:\vspace{1.5mm}\newline
{\textbf{M1:} How many websites provide cookie notices? We measure this by querying the output of the \textit{Detector} module.}\vspace{1.5mm}\newline
{\textbf{M2:} How many websites do not provide choices in the cookie notices? We query the representations of the cookie notices and count the number of websites that have only one interactable element. This set essentially includes the cookie notices which only provide one choice: to accept the cookies. }
\vspace{1.5mm}\newline
{\textbf{M3:} How many websites enable the non-essential cookies by default? We first query the cookie notice to find the number of Type A elements, \textit{i.e.} elements that can be configured. Next, we check the output of the decision model to observe if the model adjusts the preferences for any of these elements. Adjusting preferences implies that the user has to change the default settings in order to disable the non-essential cookies.}

Moreover, due to the difference in privacy regulations in Europe and the US, the websites can decide to show different content based on the geographic location of the users. To understand how this affects the cookie notices, we perform the measurement in both regions and compare the results. Specifically, we perform the measurements by accessing websites from the United Kingdom (UK) and the United States (US).

\subsubsection{Findings}
We describe the findings for the measurements between October and November 2021 using the Selenium library with ChromeDriver.

\subsubsection*{\textbf{M1: Number of cookie notices}} We measure the presence/absence of cookie notices in the top-\set websites by visiting the website and running the \textit{Detector} module (Section~\ref{sec:detector}). We find that, when accessed from the UK, we detect notices on 53\% of the websites, whereas when accessed from the US, we detect 25\% of the websites showing cookie notices. It is important to note that the websites that do not show cookie notices may still comply with the regulations if they do not store any cookies on the user browser, or if they do not use tracking cookies.
For example, \url{www.mozilla.org} does not store any cookies on the browser when accessed from both the locations. 

\iftrue %
We further note that prior work\cite{eijk2019impact}, which used 
a keyword detector based on CSS selectors 
to identify cookie notices, found that 40\% of the websites in their dataset contained cookie notices in 2019. We attribute the observed increase in cookie notices to two factors: (a) More websites have had a chance to comply with the GDPR since that time, (b) keyword-based approach can miss cookie notices which use non-standard CSS classes. We also note that their dataset was composed of the top-100 websites from 18 TLDs. 
\fi

\subsubsection*{\textbf{M2: Websites not providing choices}} Within the detected cookie notices, we now find that many websites do not provide users with choices to adjust fine-grained cookie settings. In total, 18\% of the websites containing cookie notice do not provide users with fine-grained options when accessed from the UK, whereas, when accessed from the US, the fraction is 31\%. These websites are usually websites with only one view of cookie notice with ``I Accept'' as the only button. Note that this is an important metric to measure, as compliance with regulations may require that users be given option to opt-out of non-essential cookies. We attribute the observed differences to the difference in regulations in the respective regions.

\subsubsection*{\textbf{M3: Websites enabling non-essential cookies by default}} Here, we measure the number of websites which enable non-essential cookies by default. We measure this by identifying websites where the initial state of any Type A element (elements used to enable/disable a particular cookie) is changed by the decision model. When accessed from the UK, we find that 16.7\% of the websites with cookie notices enable at least one non-essential cookie by default. In the US, this number is 22\%.  This metric is especially important as enabling non-essential cookies by default was outlawed by a recent court ruling on the basis of the ePrivacy Directive~\cite{wiedemann2020ecj, kretschmer2021cookie}. Thus, a similar analysis in the EU region could help regulators to find such violations. We also note that existing works~\cite{kampanos2021accept} relying on keyword based methods cannot reliably extract these settings due to high diversity in the text.

\section{Discussion}
\label{sec:discussion}

In this section, we discuss the deployment aspects and limitations of \name.

\subsubsection*{\textbf{Nature of Consent}} \name enables the user to automatically disable non-essential cookies. In the versions we evaluated, the user does not have to review the decision before enforcement. If informed consent per website is a requirement, we can modify the browser plugin to have a UI option which lists the summary of changes that \name will enforce (generated using the text extracted for each modified setting). 
This user interface would be similar to the one evaluated in Section~\ref{sec:eval_user_study}, with the main addition being the summary of changes.

\subsubsection*{\textbf{Impact of Design Decision}} While extracting the interactable elements in the \textit{Analyzer} module, we have made a design decision to filter out elements which take the user to a dedicated cookie settings page. We note that this decision does not impact the user experience. Take \url{www.linkedin.com} (as accessed from the UK) as an example. After the element that redirects to the cookie settings page (``Manage Preferences'')is filtered out, there are no more Type A, B or C (Table.~\ref{tab:execution_roles}) elements left on the page. Having identified this, we do not take any action on such page (thus, not clicking on ``Accept Cookies'' button) and leave the user with the cookie notice, to interact with it as they deem fit. Further, analyzing the top 200 websites manually, we find that dedicated pages for cookie settings are present in only 7 domains.

\subsubsection*{\textbf{Limitations}} One of the major limitations for \name is when a configuration to disable cookies does not exist. This entails websites which only provide one option to the user - to accept the cookies. For example, one button in the banner with the option as ``I Accept.'' Accepting such notices on users' behalf automatically might not be the desired choice. 
The UI modification that we discussed above mitigates this risk as it enables the user to decide after reviewing the summary of changes.

Another limitation for \name comes from variability in HTML implementation. For example, \name relies on an accessibility feature (\textit{tabbing}) to identify the interactive elements in the cookie notices. However, as we noted in our evaluation, the websites can implement buttons which do not fit this criteria. Empirically, we observe such websites to be rare but we accept this as a potential limitation.

Finally, we note that \name can fail during enforcement on the client side. This failure can result from change in cookie notice or the elements within it going stale. These failure modes can be detected via the plugin which can (after user consent) trigger a re-generation of the instructions by the \textit{Backend} of \name. These failures would only result in the cookie notices staying on the screen, and the user can then submit their preferences.

\section{Conclusion}
In this paper, we present \name, which, given a domain, automatically detects the cookie notice, extracts the options provided and transforms the cookie notice into a machine readable format. It then uses a text-to-text deep learning model to understand the different options provided and determines the steps required to automatically disable non-essential cookies. The machine readable format of the cookie notice further enables more usable interfaces to be built. Finally, we have extensively evaluated the performance of \name and found that it accurately annotates the cookie notices of a given domain. Further, the users also found \name's interface more usable compared to the existing baseline.
    
\newpage
\bibliographystyle{IEEEtranS}
\bibliography{references, z_ref.bib}

\appendix

\subsection{Details of the User Study}
\label{sec:appendix_baseline_study}

\subsubsection{Baseline Study}
\begin{figure}[h!]
    \centering
    \includegraphics[width=0.5\columnwidth]{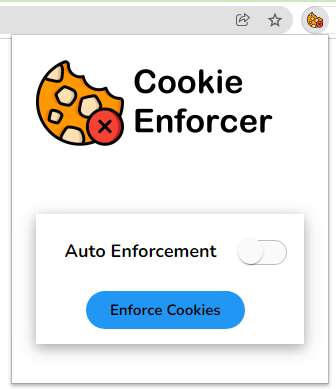}
    \caption{Chrome Browser Extension}
    \label{fig:chrm_ext}
\end{figure}

\begin{figure}[h!]
    \centering
    \includegraphics[width=\columnwidth,scale=0.9]{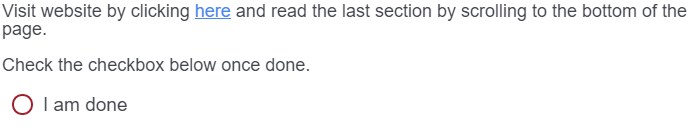}
    \caption{Prompting Users to visit the website}
    \label{fig:q_1}
\end{figure}

Here we provide more details about the User Study we conducted. We first asked users to install our custom chrome browser extension (Fig \ref{fig:chrm_ext}) which detected the websites from Table~\ref{tab:websites} which the user has never visited. From that list we choose 4 website for study. Then they were prompted to visit these website as shown in Fig. \ref{fig:q_1}.

\begin{figure}[h!]
    \centering
    \includegraphics[width=0.5\columnwidth]{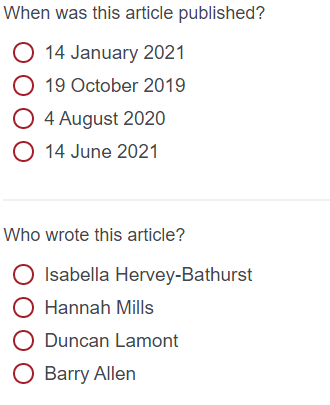}
    \caption{Prompting Users to visit the website}
    \label{fig:q_3}
\end{figure}

To make sure that the participants visited the website we asked them questions regarding the website as shown in Fig \ref{fig:q_3}.

\paragraph{Websites Used in the User Study}\label{tab:websites}

In Table~\ref{tab:websites}. we show the full list of websites used in the user study.

\begin{table}
    \begin{center}
        \begin{tabular}{
        >{\arraybackslash} m{3cm}
        >{\arraybackslash}m{3.0cm}}
        \toprule
        \textbf{Website} & \textbf{Type}                                      \\ \midrule
        \rowcolor{aliceblue} 
        harpercollins.co.uk  & Blocking   \\ \midrule
        horiba.com & Blocking \\ \midrule
        \rowcolor{aliceblue} 
        nobelbiocare.com & Blocking\\ \midrule
        vogella.com & Blocking  \\ \midrule
        \rowcolor{aliceblue} 
        schroders.com & Blocking \\ \midrule
        decathlon.co.uk & Blocking  \\ \midrule
        \rowcolor{aliceblue} 
        ledzeppelin.com & Blocking \\ \midrule
        gordonramsay.com & Non-Blocking  \\ \midrule
        \rowcolor{aliceblue} 
        crowe.com & Non-Blocking \\  \midrule
        reply.com & Non-Blocking  \\ \midrule
        \rowcolor{aliceblue} 
        quandl.com & Non-Blocking \\ \midrule
        create.arduino.cc & Non-Blocking  \\ \midrule
        \rowcolor{aliceblue} 
        financialsense.com & Non-Blocking \\
        \bottomrule
        \end{tabular}
    \end{center}
\end{table}

\paragraph{Quotes from the User Study}\label{sec:appendix_QUOTES_Baseline}
    We have manually annotated the quotes from the user study into 5 different classes: \begin{enumerate}[label=\roman*.]
      \item Forced Interaction - 
            \begin{enumerate}
                \item Typically I dismiss then but click ok if I have to to view the site.
                \item If it was blocking my view of the websites I would just click thru the cookie notice to continue with what I was doing. Other than that I didnt really pay attention to what they were saying
                \item I click X if I can typically. Or I click accept if no other option.
                \item i just approve them because the notices are cumbersome and distracting.
            \end{enumerate}
      \item Misinformed - 
            \begin{enumerate}
                \item I think cookies are necessary as it helps a system to load that particular website more fast the next time an user accesses the website.
                \item I just assume that cookies are part of every site so click the button to make it disappear whenever I see it or else it takes up too much of the page.
                \item It is necessary to to allow cookies in order to be able to access the site fully
                \item Since allowing cookie, will give my more effective result of the website, i just go ahead and accept it
            \end{enumerate}
      \item Lawful/Trustworthy - 
            \begin{enumerate}
                \item I typically try to read the data-sharing policy if I have time, or I base my decisions with respect to cookies on my overall trust in the website/company.
                \item just intially recognized the website is a normal one without any protential risk.
                \item I typically allow cookies for any website that appears legitimate and respected. 
                \item If the site seems trustworthy I feel better about it.
            \end{enumerate}
      \item Risk Dependent - 
            \begin{enumerate}
                \item I have heard about tracking cookies so I am weary every time I choose to accept cookies.
                \item I just worry I'm being tracked by my cookies. So I prefer to not allow cookies
                \item I prefer not to allow them but I delete them from my browser nightly regardless.
                \item I don't allow cookies if I have a choice.
            \end{enumerate}
      \item Others - 
            \begin{enumerate}
                \item I always accept cookies and always click the notices. Actually I wish I could find an add-on or such that would click them all automatically. I think it indicates it is a British site when you see these (perhaps other countries as well). I don't mind cookies in most cases.
                \item I don't really think about it.
                \item Nothing will affect my cookie settings.  I just accept and continue cookies.
                \item I just gave allowed to cookie settings.
            \end{enumerate}
    \end{enumerate}

    \subsubsection{Usability Study}
\begin{figure}[h!]
    \centering
    \includegraphics[width=\columnwidth,scale=0.9]{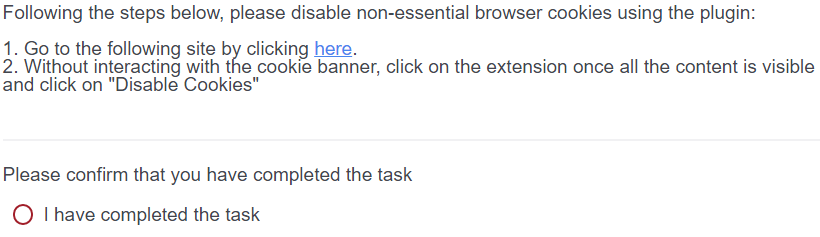}
    \caption{Tasks users have to complete on the website linked}
    \label{fig:task}
\end{figure}

    Here we provide more details about the User Study we conducted. We first asked users to install our custom chrome browser extension (Fig \ref{fig:chrm_ext}) which detected the websites from Table \ref{tab:websites} which the user has never visited. From that list we choose 2 website for study. Then they were prompted to complete the tasks on the websites as shown in Fig. \ref{fig:task}.
    
\begin{figure}[h!]
    \centering
    \includegraphics[width=\columnwidth]{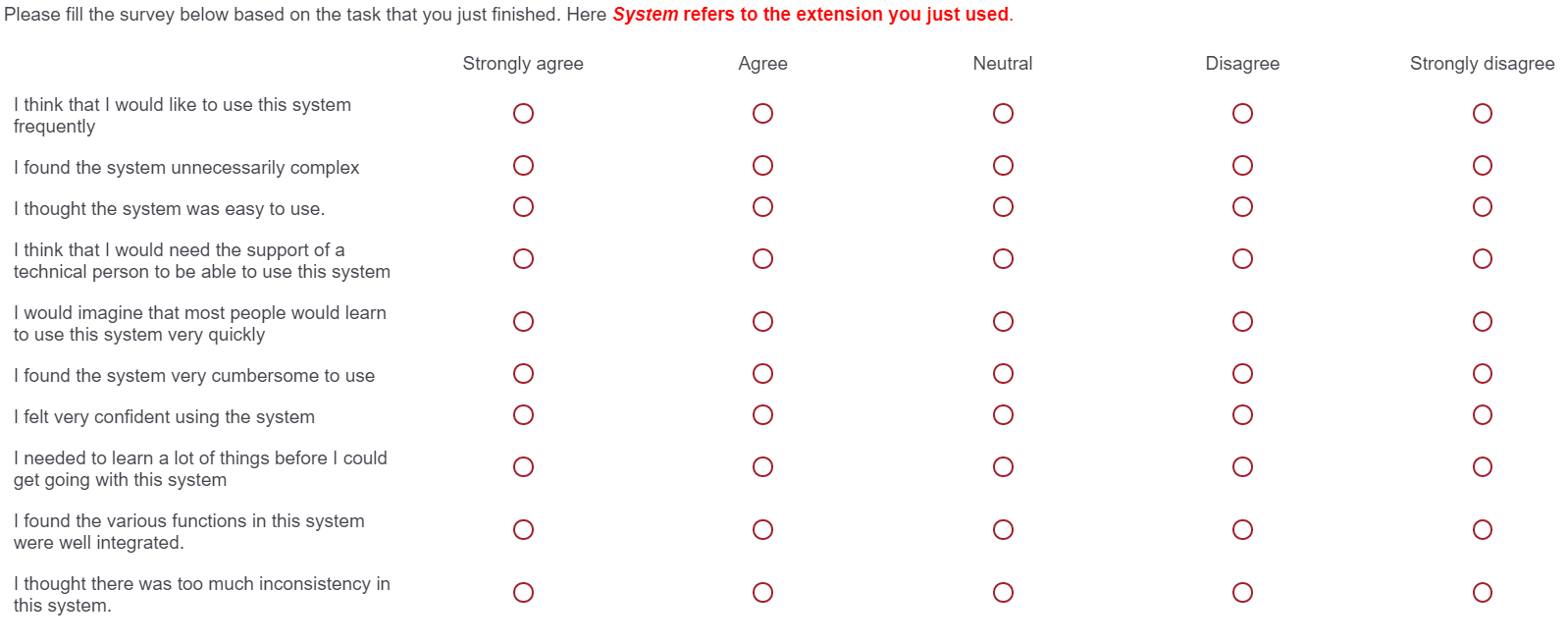}
    \caption{System Usability Scale questionnaire}
    \label{fig:sus_q}
\end{figure}

    \paragraph{Usability Evaluation}
At the end of the User Study we asked the participants to fill out the System Usability Scale questionnaire \cite{brooke1996sus}. A snapshot of the questionaire is shown in Fig. \ref{fig:sus_q}.

\subsection{Examples of Cookie Notices}
\label{sec:appendix_t5_examples}
Here, we show some examples of the cookie banners that are discussed in Section~\ref{sec:decision}.
\begin{figure*}[t]
    \centering
    \includegraphics[width=\textwidth]{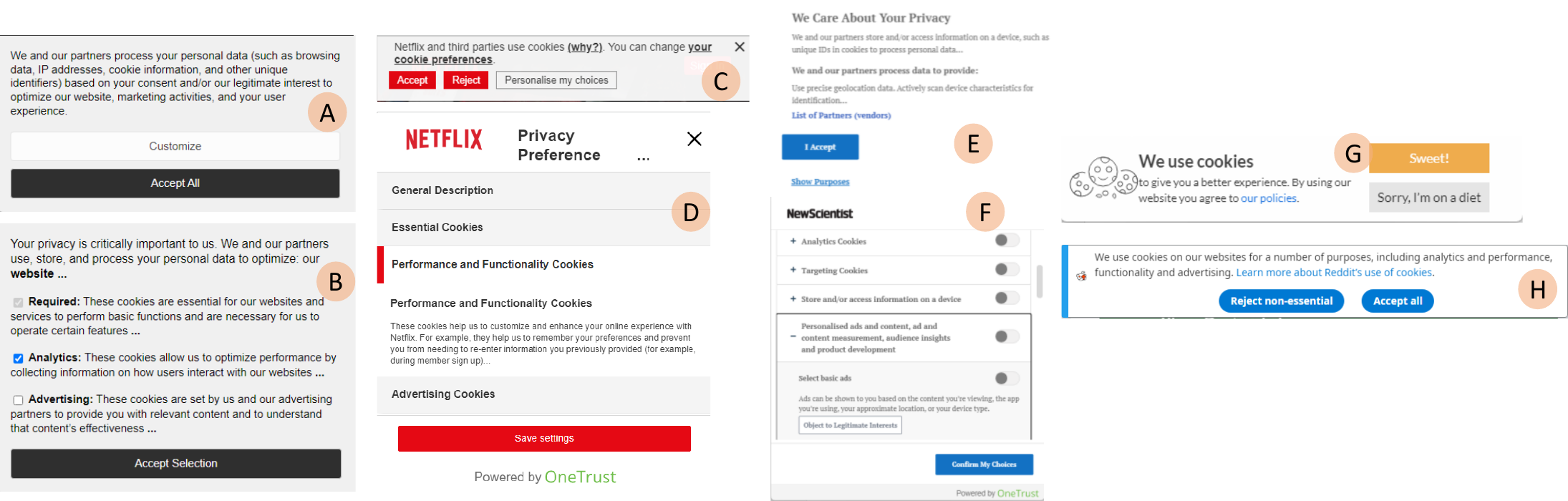}
    \caption{Screenshots of Cookie Notices for the websites listed in Table \ref{table:t5_examples} where (A) is the first view for WordPress and (B) is the second view. (C) is the first view for Netflix and (D) is the second view. (E) is the first view of NewScientist and (F) is the second view .(G) is the first view of TATA and (H) is the first view of Reddit}
    \label{fig:cookie_examples}
\end{figure*}

\end{document}